\documentclass[letterpaper,twocolumn,10pt]{article}
\usepackage{USENIX}

\usepackage{tikz}
\usepackage{amsmath}

\usepackage{filecontents}

\usepackage{graphicx}
\usepackage{subfigure}
\usepackage{amsmath}
\usepackage{amsthm}
\usepackage{mathrsfs}
\usepackage[vlined,ruled,linesnumbered,algo2e]{algorithm2e}
\usepackage{xcolor}
\usepackage{multicol}
\usepackage{multirow}
\usepackage{diagbox}
\usepackage{bbm}
\usepackage{makecell}
\usepackage{enumitem}
\usepackage{booktabs}
\usepackage{algorithm}
\usepackage{algorithmic}
\usepackage{appendix}
\usepackage{hyperref}
\usepackage{inconsolata}

\begin{document}

\date{}

\title{Data-Free Model-Related Attacks: Unleashing the Potential of Generative AI}


\author{{\rm Dayong Ye$^*$} \hspace{2mm} {\rm Tianqing Zhu$^{\dagger}$} \hspace{2mm} {\rm Shang Wang$^*$} \hspace{2mm} {\rm Bo Liu$^*$} \\ {\rm Leo Yu Zhang$^{\P}$} \hspace{2mm} {\rm Wanlei Zhou$^{\dagger}$} \hspace{2mm} {\rm Yang Zhang$^{\S}$}\\
$^*$University of Technology Sydney \hspace{2mm} $^{\dagger}$City University of Macau \\ $^{\P}$Griffith University \hspace{2mm} $^{\S}$CISPA Helmholtz Center for Information Security}

\maketitle

\renewcommand{\thefootnote}{}
\footnotetext{Tianqing Zhu is the corresponding author.}

\begin{abstract}
Generative AI technology has become increasingly integrated into our daily lives, offering powerful capabilities to enhance productivity. However, these same capabilities can be exploited by adversaries for malicious purposes. While existing research on adversarial applications of generative AI predominantly focuses on cyberattacks, less attention has been given to attacks targeting deep learning models. In this paper, we introduce the use of generative AI for facilitating model-related attacks, including model extraction, membership inference, and model inversion. Our study reveals that adversaries can launch a variety of model-related attacks against both image and text models in a data-free and black-box manner, achieving comparable performance to baseline methods that have access to the target models' training data and parameters in a white-box manner. This research serves as an important early warning to the community about the potential risks associated with generative AI-powered attacks on deep learning models. The source code is provided at: https://zenodo.org/records/14737003. 
\end{abstract}

\maketitle

\section{Introduction}
Generative AI has demonstrated its potent capabilities in both image and language processing recently \cite{GM20,OpenAI22DALLE,openai2023GPT4}. Along with their broad availability, concerns regarding the privacy and security implications associated with their usage have also emerged \cite{Gupta23}. These concerns can be roughly classified into two categories \cite{Yao23}: 1) offensive applications, involving the use of generative AI for malicious purposes \cite{Chen24ICLR}; and 2) potential vulnerabilities, referring to weaknesses that can be exploited to compromise a generative model's own privacy, such as extracting their training data \cite{Carlini21USENIX,Li23EMNLP}. Among these two categories, current research primarily focuses on the second category, while the first category, i.e., the offensive applications of generative AI, has received limited attention. 

Current research on the offensive applications of generative AI primarily targets cyberattacks, such as FraudGPT \cite{FraudGPT}. However, the potential of generative AI to facilitate model-related attacks, such as model extraction \cite{Oliynyk23ACMCUR}, membership inference \cite{Shokri17,Ye22}, and model inversion attacks \cite{Yang19CCS,Zhao21ICCV,Zhu23TIFS}, remains largely underexplored.
While model-related attacks are well-documented in the context of existing machine learning models, they typically operate under the assumption that the adversary has access to a dataset that shares the same or similar distribution as the target model's training set \cite{Choquette21ICML,Kahla22CVPR,Oliynyk23ACMCUR}. However, this assumption is often impractical in real-world scenarios due to various constraints such as the limited capability of the adversary to access proprietary datasets. While some studies have explored attacks in a data-free manner, they often depend on having white-box access to the target model, meaning the adversary possesses detailed knowledge of the model's parameters and architecture \cite{Liu22USENIX}. Alternatively, they lacking such access may experience a decrease in attack performance \cite{Mattern23ACL}. Another research trend is using large language models for prompt stealing, wherein prompts are reconstructed based on the corresponding responses \cite{Sha24,Yang24}. Our research differs significantly in two key aspects. Firstly, our study delves into various model-related attacks, while prompt stealing research primarily focuses on this singular objective. Secondly, our approach is conducted in a data-free manner, whereas prompt stealing methodologies typically rely on a substantial amount of externally collected data.

In this paper, we embark on a pioneering exploration of the offensive applications of generative AI. 
The novelty of this work lies in leveraging generative AI models to conduct various model-related attacks across both image and text domains in a data-free and black-box manner. Unlike existing research that typically relies on external datasets to launch attacks, generative models eliminate the need for such data by directly generating high-quality synthetic data. This significantly lowers the barrier for executing model-related attacks. The significance of this work, therefore, lies in exposing the potential security risks associated with the misuse of generative AI models.
However, soliciting the generative model to generate effective data poses challenges due to two reasons. 

\begin{itemize}[leftmargin=*]

    \item Generating data that meets the diverse requirements of various model-related attacks and comprehensively covers the sample space, while preserving essential characteristics, poses a challenge. Depending on the nature of the attack, the generated samples must exhibit specific characteristics tailored to the attack's objectives. For instance, in membership inference attacks, data samples proximate to decision boundaries are crucial, while in model inversion attacks, samples representing distinct classes are necessary. 
    
    \item Mitigating the distribution shift between the target model's training data and the generated data presents a challenge. This shift can arise from the inherent randomness in the generation process of generative models, potentially leading to discrepancies in data characteristics. Such discrepancies can degrade the performance of the generated data when employed in tasks like model-related attacks, where precise data characteristics are crucial for effectiveness.
\end{itemize}

To tackle these challenges, we present a novel data generation approach. Leveraging insights into the target model's task, the adversary meticulously designs prompts to direct the generative model in generating the required data. Subsequently, the adversary employs data augmentation techniques to diversify the generated samples.  In particular, the augmentation process involves exploring the decision boundary of the target model and collecting samples in its vicinity. To further address the distribution shift, we introduce an inter-class filtering approach. This approach filters out anomalous samples by comparing the distances of generated samples to the class centroids in the feature space defined by the target model's outputs. 
In summary, this work makes four contributions.

\begin{itemize}[leftmargin=*]
    \item This paper identifies a significant vulnerability in deep learning models, namely their susceptibility to model-related attacks enabled by generative AI. It offers the first comprehensive and generalized study exploring the potential of generative AI to execute model-related attacks across three prevalent types, all without using externally collected data. 


    \item We propose a novel data generation approach utilitizing the potent capabilities of generative AI. This approach is tailored to generate near-boundary samples, effectively spanning the entirety of the sample space. 

    \item We introduce an innovative inter-class filtering approach designed to mitigate the distribution shift between the target model's training data and the generated data. This method significantly enhances the quality and usability of generated datasets for model-related attacks.
    
    \item We undertake comprehensive experiments to assess the efficacy of our method. Unlike the majority of studies in model-related attacks that solely rely on image data, our evaluation encompasses both image and text data. 
\end{itemize}

\section{Preliminary and Threat Model}

\noindent\textbf{Image-based Generative Models.} 
This work adopts diffusion models \cite{Yang23ACMCUR} for its generative tasks due to its superior performance compared to VAEs \cite{Kingma19} and GANs \cite{Goodfellow14}.

A diffusion model is typically characterized by a forward process that introduces noise to data and a reverse process that reverts noise back to data. The forward process serves to convert any data distribution into a straightforward prior distribution, such as a standard Gaussian, while the reverse process involves learning transition kernels parameterized by deep neural networks to invert the forward process. Hence, new data samples can be generated by initially sampling a random vector from the prior distribution and then performing ancestral sampling through the reverse process.


\vspace{2mm}
\noindent\textbf{Language-based Generative Models.} One of the most influential language-based generative models is large language models (LLMs) exemplified by GPT-4 \cite{openai2023GPT4}.
A large language model, denoted as $LLM$, functions as a transformative tool in natural language understanding and generation. It takes a text sample $x$, often referred to as a `prompt', as input and generates another text sample $y$, typically referred to as a `response', i.e., $y=LLM(x)$. 
Text samples processed by LLMs are represented as sequences of tokens, where $x = [x^1, x^2, \cdots, x^s]$, and $s$ represents the number of tokens in $x$. These tokens are discrete units of language, which could be words, subwords, or even characters, depending on the tokenization scheme used. 

\vspace{2mm}
\noindent\textbf{Threat Model.} Given a target model $T$, the adversary has knowledge about the task for which the model $T$ is designed, i.e., the adversary knows the meanings associated with each class $c$ of the target model $T$. This assumption is reasonable, as model providers typically make this information public whenever the model is used to offer services to the public. 

The adversary interacts with $T$ in a black-box manner, meaning they can query $T$ and observe its output confidence vectors, but they lack access to $T$'s parameters and architecture. Note that our method can also be adapted to scenarios where the adversary can only observe the output hard labels of $T$, without access to confidence scores. 

The adversary is capable of locating a black-box generative model $F$ designed to perform a task closely aligned with that of the target model. For instance, in the case where the target model serves as an image classifier, the adversary is able to identify an image-based generative model. Once this generative model $F$ is obtained, the adversary has the capacity to submit queries to it. 
The goal of the adversary is to construct a dataset, denoted as $\mathcal{D}_{aux}$, which shares the same or similar distribution as the dataset, $\mathcal{D}_{train}$, used to train the target model $T$. 
After acquiring $\mathcal{D}_{aux}$, the adversary can conduct a range of model-related attacks. 
The specifics of these attacks are outlined below.

\vspace{2mm}
\noindent\textbf{Model Extraction.} The objective of model extraction attacks, also referred to as model stealing, is to create a model $E$ that is functionally identical or closely resembles the target model $T$ \cite{Oliynyk23ACMCUR}. 
Formally, for a given input example $x$, the adversary aims for $E(x) = T(x)$. 

\vspace{2mm}
\noindent\textbf{Membership Inference.} The objective of membership inference attacks is to ascertain whether a given example $x$ belongs to the training set $\mathcal{D}_{train}$ of the target model $T$ \cite{Shokri17,Mattern23ACL}. In executing membership inference, the adversary commonly trains a binary classification model $A$. This model takes either the final output $T(x)$ or the intermediate layers' output of the target model — depending on the adversary's level of information — as input and produces a scalar in the range $[0,1].$ This scalar denotes the probability that $x$ is in $\mathcal{D}_{train}$. 


\vspace{2mm}
\noindent\textbf{Model Inversion.} Model inversion seeks to reconstruct data samples from a target model $T$ by observing its output. 
Depending on the nature of the target model $T$'s output, model inversion can be categorized into two types. The first type aims to reconstruct a representative sample for each class of the target model $T$ \cite{Kahla22CVPR,Zhu23TIFS,Han23CVPR}, particularly when $T$ outputs only hard labels. Formally, for each class $c$ of the target model $T$, the adversary aims to generate an example $\hat{x}$ that captures the essential features of the samples in class $c$. The second type seeks to reconstruct any input samples presented to the target model $T$ \cite{Yang19CCS,Zhao21ICCV}, especially when $T$ outputs confidence vectors. Formally, when given an input example $x$, the adversary aims to reconstruct $x$ by leveraging $T(x)$.

\section{Methodology}
\subsection{Overview}
The rationale behind the proposed method stems from the common practice of training generative models on extensive datasets. For instance, the development of DALL-E, publicly introduced alongside with CLIP (Contrastive Language-Image Pre-training), involved training on a substantial dataset consisting of $400$ million pairs of images with text captions extracted from the Internet \cite{Johnson21}. Consequently, the distribution space of training data for generative models is highly likely to encompass the distribution space of the training set for the target model if the target model shares a similar task with the generative model. Therefore, the primary task for the adversary is to uncover the distribution of the training set associated with the target model. The general attack pipeline operates as follows. First, the adversary queries the generative model to produce data, which is subsequently used to query the target model. The outputs collected from the target model provide insights into the distribution of its training set and are then leveraged to execute model-related attacks.


Specifically, our method comprises \textbf{three steps}. \textbf{Firstly}, leveraging their understanding of the task of the target model $T$, the adversary identifies a publicly available generative model $F$ that closely aligns with $T$. For instance, in the case of an image classifier target model, the adversary might opt for DALL-E \cite{OpenAI22DALLE} as the generative model. Then, armed with knowledge about the meanings of each class $c$ within $T$, the adversary instructs the generative model $F$ to generate a set of samples for each class $c$, denoted as $\mathcal{D}_c$. By uniting these $\mathcal{D}_c$ sets across all classes, the adversary compiles a dataset $\mathcal{D}_a=\bigcup^C_{c=1}\mathcal{D}_c$, where $C$ represents the number of classes. 

\textbf{Secondly}, the generated dataset $\mathcal{D}_a$, however, may not inherently share the same distribution as model $T$'s training set $\mathcal{D}_{train}$. This discrepancy arises from potential distinctive features present in synthetic samples generated by the model $F$, as opposed to the often indistinct features characterizing real samples in $\mathcal{D}_{train}$. To address this distinction, the adversary augments $\mathcal{D}_a$ to $\mathcal{D}_{aux}$ by generating additional synthetic samples with indistinct features. 

\textbf{Finally}, the adversary implements inter-class filtering to remove the outlying samples from the generated dataset $\mathcal{D}_{aux}$. The refined dataset resulting from this process is denoted as $\widehat{\mathcal{D}_{aux}}$. Then, the adversary employs $\widehat{\mathcal{D}_{aux}}$ to train a model $\hat{T}$ with the objective of mimicking the functionality of $T$.

\subsection{Details of the Method}
We formally describe the three steps in detail as follows.

\vspace{2mm}
\noindent\textbf{Step 1: Generate data samples using a generative model.}
To produce samples within class $c$, the adversary can straightforwardly employ natural language to guide the generative model $F$. For example, by utilizing a prompt like ``please generate $n$ samples that exhibit key features of class $c$'', within GPT-4.0, the adversary instructs DALL-E to generate samples accordingly. To assess the efficacy of this set of generated samples, denoted as $\mathcal{D}_c$, the adversary uses them to query the target model $T$. Any samples not correctly classified by $T$ into class $c$ are subsequently discarded.

\vspace{2mm}
\noindent\textbf{Step 2: Enhance the generated data through augmentation.}
Following the generation of samples across all classes, the adversary acquires a consolidated set $\mathcal{D}_a$. Nevertheless, as all samples within $\mathcal{D}_a$ are generated by the model $F$, they exhibit distinctive features. Consequently, directly employing $\mathcal{D}_a$ to train a model $\hat{T}$ would lead to a degradation in model performance compared to the target model $T$, as shown in the experimental section. Therefore, the adversary must augment $\mathcal{D}_a$ by introducing additional samples with indistinct features.

\begin{figure}[ht]
\centering
	\includegraphics[scale=0.4]{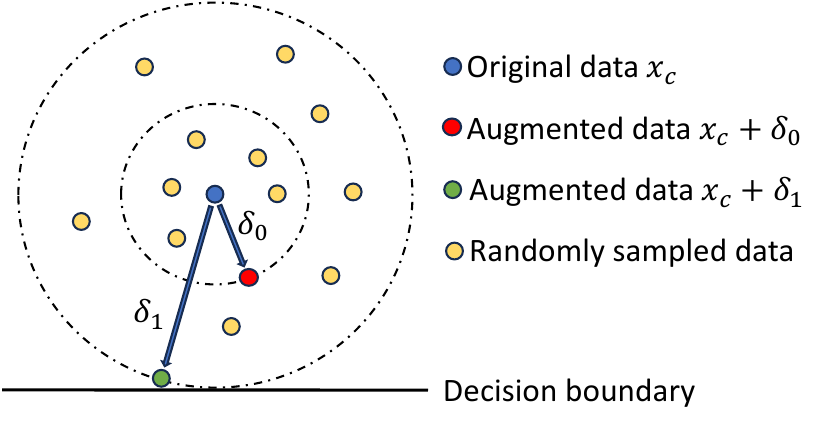}
	\caption{Exploration of the decision boundary in the target model. Noise is incrementally added until the boundary is crossed, leading to misclassification by the target model.} 
	\label{fig:boundary}
\end{figure}

To augment each $\mathcal{D}_c$ within $\mathcal{D}_a$, the adversary begins by randomly selecting a sample $x_c$ from $\mathcal{D}_c$. Since $x_c$ is correctly classified by the target model $T$ into class $c$, the adversary introduces a small noise $\delta_0$ to $x_c$, i.e., $x_c + \delta_0$, and gradually increases the noise amount to explore the decision boundary of $T$, as shown in Figure \ref{fig:boundary}. $\delta_0$ represents the initial amount of Gaussian noise, serving as the starting point for augmentation. With each increase from $\delta_{i-1}$ to $\delta_i$, the adversary randomly collects $N$ samples within a sphere, using $x_c$ as the centroid with a radius ranging from $\delta_{i-1}$ to $\delta_i$. If all collected samples are correctly classified as class $c$ by the target model $T$, the adversary proceeds to increase the noise from $\delta_i$ to $\delta_{i+1}$ and repeats the process. However, if the noise is large enough to cross the decision boundary, the adversary concludes the process, and considers $\delta_i$ as the potential maximum noise amount.
Moreover, all previously collected samples, employed to assess the decision boundary of $T$, are incorporated to augment $\mathcal{D}_a$. This process is formally detailed in Algorithm \ref{alg:augmentation}.

\begin{algorithm}\small
\caption{The data augmentation approach}
\label{alg:augmentation}
\begin{algorithmic}[1]
\REQUIRE{A set of $C$ generated datasets: $\mathcal{D}_1, \cdots,\mathcal{D}_C$; The initial noise $\delta_0$ and the noise step size $\epsilon$;}
\ENSURE{The augmented auxiliary dataset $\mathcal{D}_{aux}$;}
\FOR{each class $c$}
    \STATE Randomly select a sample $x_c$ from $\mathcal{D}_c$;
    \FOR{each test round $i=0,1,2, \cdots$}
        \STATE Randomly collect $N$ samples within a space, using $x_c$ as the centroid with a radius ranging between $\delta_{i-1}$ and $\delta_i$;
        \IF{for any sample $x_j$, $j=1,\cdots,N$, that $T(x_j)\neq c$}
            \STATE \textbf{break;}
        \ENDIF
        \STATE $\mathcal{D}_c\leftarrow\mathcal{D}_c\bigcup\{x_1,...,x_N\}$;
        \STATE $\delta_{i+1}\leftarrow\delta_i+\epsilon$;
    \ENDFOR
\ENDFOR
\RETURN $\mathcal{D}_{aux}\leftarrow\bigcup^C_{c=1}\mathcal{D}_c$;
\end{algorithmic}
\end{algorithm}

Within Algorithm \ref{alg:augmentation}, the initial noise $\delta_0$ and the noise step size $\epsilon$ stand as hyperparameters requiring fine-tuning. A large $\delta_0$ may cause the algorithm to halt in the initial test round without any augmentation, i.e., $x_c+\delta_0$ has already traversed the decision boundary of the target model $T$. Similarly, a large $\epsilon$ can render it challenging for the adversary to precisely pinpoint the decision boundary. 

In Line 4, the collection of $N$ samples is achieved by introducing random noise to $x_c$, with the noise magnitude restricted to $\delta_i$. It is worth noting that if the noise $\delta_i$ is represented as a vector, its magnitude can be confined by utilizing an $L_p$ norm. This norm constraint ensures that the noise does not exceed a predefined limit, contributing to controlled perturbations in the sample space.

\vspace{2mm}
\noindent\textbf{Step 3: Distribution Shift Mitigation.} After the data generation and augmentation steps, the resulting dataset $\mathcal{D}_{aux}$ may differ in distribution from the target model $T$'s training set due to randomness introduced during these steps. Using this distribution-shifted data as a training set may not yield a model functionally similar to the target model. To address this issue, we employ inter-class information to filter $\mathcal{D}_{aux}$.

To perform inter-class filtering, we begin by feeding the generated samples from each class $\mathcal{D}_i$ into the target model $T$ and collecting the corresponding output vectors, where $i\in[1,C]$ and $C$ denotes the number of classes. Formally, let $x^i_j$ represent a sample in $\mathcal{D}_i$. The output vectors are then collected as $T(x^i_1),...,T(x^i_{n_i})$, where $n_i$ denotes the number of samples in $\mathcal{D}_i$. The centroid of the samples in $\mathcal{D}_i$ is computed by averaging these output vectors: $Cen_i=\frac{1}{n_i}\sum^{n_i}_{j=1}T(x^k_j)$.
Notably, we compute the centroid using the samples' corresponding output vectors from the target model $T$, rather than the samples themselves, for two key reasons. First, the output vectors of the target model encapsulate the model's understanding and internal representation of the data. Calculating the centroid in this transformed space aligns the data analysis with the target model's learned features and structure, which is crucial for effectively mimicking the target model. Second, the output vectors from the target model typically have much lower dimensionality compared to the raw data. This dimensionality reduction simplifies computations and reduces noise, resulting in a more robust and efficient centroid calculation.

After computing the centroid for the generated samples from each class $\mathcal{D}_i$, we calculate the distance between the centroid of one class and the samples from each of the other classes. Formally, let $Cen_i$ represent the centroid of class $i$, and $k$ denote any other class. The generated samples in class $k$ are represented as $\mathcal{D}_k=\{x^k_1,...,x^k_{n_k}\}$. By feeding $x^k_1,...,x^k_{n_k}$ into the target model $T$, we obtain the corresponding output vectors: $T(x^k_1),...,T(x^k_{n_k})$. The distance between $Cen_i$ and each $T(x^k_j)$, where $j\in[1,n_k]$, is calculated using the $l_2$ norm: $||Cen_i - T(x^k_j)||_2$. Once the distances are computed, we apply the three-sigma rule to filter out those samples in class $k$ that fall beyond three standard deviations from the mean of the distribution. 
The rationale is grounded in the behavior of the target model $T$, trained on the original data set, $\mathcal{D}_{train}$. The output vectors from $T$ reflect the learned distribution characteristics of $\mathcal{D}_{train}$. By filtering out samples whose output vectors are statistical outliers or do not conform to the expected class distribution, captured by the centroids, the distributional shifts between $\mathcal{D}_{aux}$ and $\mathcal{D}_{train}$ can be minimized. 
This filtering process is summarized in Algorithm \ref{alg:filtering}. 

\begin{algorithm}\small
\caption{The inter-class filtering approach}
\label{alg:filtering}
\begin{algorithmic}[1]
\REQUIRE{The generated auxiliary dataset $\mathcal{D}_{aux}$;}
\ENSURE{The filtered auxiliary dataset $\widehat{\mathcal{D}_{aux}}$;}
\FOR{each class $i$ in $\mathcal{D}_{aux}$}
    \FOR{each sample $x^i_j$ in class $i$}
        \STATE Feed $x^i_j$ into the target model $T$ and receive the output vector $T(x^i_j)$;
    \ENDFOR
    \STATE Compute the centroid of the samples in class $i$ as: $Cen_i=\frac{1}{n_i}\sum^{n_i}_{j=1}T(x^i_j)$, where $n_i$ denotes the number of samples;
    \FOR{each of the other classes, denoted as $k$, in $\mathcal{D}_{aux}$}
        \FOR{each sample $x^k_j$ in class $k$}
            \STATE Compute the distance between the centroid $Cen_i$ and $x^k_j$ as: $||Cen_i - T(x^k_j)||_2$;
        \ENDFOR
        \STATE Apply the three-sigma rule to filter out those samples in class $k$ that fall beyond three standard deviations from the mean of the distribution;
    \ENDFOR 
\ENDFOR
\RETURN $\widehat{\mathcal{D}_{aux}}$;
\end{algorithmic}
\end{algorithm}


\subsection{Conduct Model-Related Attacks}
After obtaining the generated dataset $\widehat{\mathcal{D}_{aux}}$, the adversary can initiate a series of model-related attacks.

\vspace{2mm}
\noindent\textbf{Model Extraction Attack.} The aim of a model extraction attack is to obtain a model that is functionally equivalent to the target model $T$. To achieve this, the adversary ensures that the mimicking model $\hat{T}$ has a similar classification accuracy to $T$ on a given test set. 
The adversary splits the dataset $\widehat{\mathcal{D}_{aux}}$ into two subsets: a training set $\widehat{\mathcal{D}^{train}_{aux}}$ and a test set $\widehat{\mathcal{D}^{test}_{aux}}$. Subsequently, $\hat{T}$ is trained on $\widehat{\mathcal{D}^{train}_{aux}}$, and both $T$ and $\hat{T}$ are evaluated on $\widehat{\mathcal{D}^{test}_{aux}}$. If $\hat{T}$ exhibits lower accuracy than $T$, the adversary may adjust hyperparameters, explore different architectures, and request the generative model to generate additional training data to enhance $\hat{T}$'s performance.

Note that the dataset $\widehat{\mathcal{D}_{aux}}$ is constructed based on the filtering process of the target model $T$, where the adversary leverages $T$ to classify all generated data. Consequently, $\widehat{\mathcal{D}_{aux}}$ inherently retains traces of $T$ because misclassifications by $T$ can introduce biases into $\widehat{\mathcal{D}_{aux}}$. For instance, a generated sample $x$ with a ground truth label $y$ may be misclassified by $T$ as $y'$, due to the imperfectness of $T$. Thus, $x$ would not be included in $\widehat{\mathcal{D}_{aux}}$, and this bias, incurred by $T$, is carried by $\widehat{\mathcal{D}_{aux}}$. While biases themselves are undesirable, they can be strategically exploited by the adversary to efficiently mimic the behavior of the target model $T$.
Specifically, the biases in $\widehat{\mathcal{D}_{aux}}$ align with the decision-making tendencies of $T$. By exploiting these biases, the adversary can efficiently train the mimicking model $\hat{T}$ to replicate $T$'s behavior without needing to explicitly model complex decision boundaries.

\textbf{The label-only scenario} is equivalent to the vector-based scenario described above. This is because, in the model extraction attack, the adversary's goal is to match the test accuracy of the target model $T$. To achieve this, the adversary relies solely on the hard label outputs of the target model $T$. As a result, the attack process remains consistent, regardless of whether the adversary has access to confidence vectors or operates in a label-only setting.

\vspace{2mm}
\noindent\textbf{Membership Inference Attack.} To conduct membership inference against the target model $T$, the adversary utilizes the mimicking model $\hat{T}$ as a shadow model to train an attack model $A$. This approach leverages the functional similarity between $\hat{T}$ and $T$, enabling $A$ to exploit patterns learned by $\hat{T}$ to infer membership status. During the training of the attack model, the adversary inputs samples from both the training set $\widehat{\mathcal{D}^{train}_{aux}}$ and the test set $\widehat{\mathcal{D}^{test}_{aux}}$ into $\hat{T}$ and collects the corresponding confidence vectors. These confidence vectors, along with the true label of each sample, are combined with a membership label (either $in$ or $out$) to create training samples for the attack model $A$. Specifically, if a sample belongs to the training set, the corresponding training sample for $A$ is $((\hat{T}(x),y),in)$; otherwise, it is $((\hat{T}(x),y),out)$.

The key insight here is that the attack model $A$, trained based on the outputs of $\hat{T}$, effectively learns to discern between members (i.e., samples from the training set of $\hat{T}$) and non-members (i.e., samples not from the training set of $\hat{T}$). Since $\hat{T}$ is functionally similar to $T$, the attack model $A$ can also exploit similar patterns and characteristics to discern membership in $T$. This implies that $A$, trained using the mimicking model $\hat{T}$, can be directly applied to conduct membership inference attacks against the target model $T$.

In \textbf{the label-only scenario}, the attack process simplifies significantly. Specifically, Step 2 of our method can be directly applied to conduct the membership inference attack solely based on labels. Here, the amount of noise needed to push a data sample across the decision boundary of the target model serves as a proxy for the model's confidence in predicting that sample's label \cite{Li21CCS}. Larger noise amounts indicate higher confidence levels. Thus, if the noise surpasses a certain threshold, the data sample can be deemed a member of the target model's training set. 
This approach uses the concept of noise sensitivity as a proxy for data familiarity. The rationale is that data points that are members of the training set will typically be more robust to perturbations (noise), while non-members are more likely to be misclassified with even slight modifications. Thus, for any given sample, if the amount of noise required to induce a misclassification exceeds a predetermined threshold, the sample is deemed as a member.

\vspace{2mm}
\noindent\textbf{Model Inversion Attack.} To execute a model inversion attack against the target model $T$, the adversary uses the generated dataset $\widehat{\mathcal{D}_{aux}}$ to train an inversion model $I$. Specifically, each data sample $x$ from $\widehat{\mathcal{D}_{aux}}$ is fed into the target model $T$, yielding the output confidence vector $T(x)$. Pairing each $x$ with its corresponding $T(x)$, denoted as $(T(x),x)$, the adversary compiles a set of training samples for training the inversion model $I$. This inversion model takes confidence vectors as input and generates an image (if $T$ is an image-based model) or a prompt (if $T$ is language-based). For the attack, given a confidence vector $T(x^*)$, the adversary can reconstruct $x^*$ by directly inputting $T(x^*)$ into $I$ and observing its output.

In \textbf{the label-only scenario}, the inversion process simplifies significantly. When given an output label, the goal shifts to reconstructing a representative input rather than precisely recovering the exact input. Precisely recovering the exact input is challenging due to the fact that a single output label may correspond to multiple distinct inputs. For instance, consider the output label ``dog''. There could be countless variations of dog images that fall under this category. Thus, in a label-only model inversion attack, the adversary's objective is to randomly select a data sample from each class in $\widehat{\mathcal{D}_{aux}}$ and utilize it as the representative input for that class. 
Note that even if the label-only model inversion attack only provides examples of a given class, these examples may inadvertently reveal sensitive characteristics about the individuals represented in the training data. For instance, if a model is trained to classify medical images and the attack produces typical examples of a disease class, the generated images might depict a rare medical condition. If it is known that a small group of individuals has been treated for this condition in a certain area, it might be possible to infer the identity of those individuals.

\section{Analyses of the Method}\label{sec:analyses}
The analysis focuses on the data distribution shifts caused by generative models, examining how these models induce shifts in data distribution and the measures taken to mitigate them.

\vspace{2mm}
\noindent\textbf{Distribution Shift Analysis.} Our analysis primarily focuses on diffusion models, but the principles can be extended to LLMs, as both utilize an iterative refinement process that introduces uncertainty or randomness. This randomness is a critical factor in the potential distribution shifts observed in both types of models. 
We begin by identifying the origins of the distribution shift and then explain how the proposed method effectively mitigates this shift.

The operation of a diffusion model is divided into two main phases: the forward process and the reverse process. In the forward process, Gaussian noise is incrementally added to the data across multiple steps, progressively transforming it into pure noise. This process is mathematically represented as: 
\begin{equation}\nonumber
    q(x_t|x_{t-1})=\mathcal{N}(x_t;\sqrt{\alpha_t}x_{t-1},(1-\alpha_t)\mathbf{I}),
\end{equation}
where $\alpha_t$ is a variance schedule that controls the amount of noise added at each step, $x_t$ represents the noisy data at time step $t$, and $\mathbf{I}$ is the unit vector. In contrast, the reverse process aims to denoise the data incrementally. Given the data $x_t$, the model predicts the data from the previous step, $x_{t-1}$, as: 
\begin{equation}\nonumber
    p_{\theta}(x_{t-1}|x_t)=\mathcal{N}(x_{t-1};\mu_{\theta}(x_t,t,z),\sigma^2_t\mathbf{I}),
\end{equation}
where $\mu_{\theta}$ and $\sigma^2_t$ are the mean and variance predicted by the model, respectively, and $z$ represents the embedding encoded from the input prompt by an encoder $E$, such as a transformer.

Specifically, the reverse process aims to reconstruct $x_{t-1}$ from $x_t$. This is done by removing the predicted noise component $\epsilon_{\theta}(x_t,t,z)$ from $x_t$ and adding some uncertainty to account for the stochastic nature of the process \cite{Ho20NIPS,Saharia22NIPS}:
\begin{equation}\label{eq:reverse}
    x_{t-1}=\frac{1}{\sqrt{\alpha_t}}(x_t-\frac{1-\alpha_t}{\sqrt{1-\overline{\alpha_t}}}\epsilon_{\theta}(x_t),t,z)+\sigma_t\xi,
\end{equation}
where $\overline{\alpha_t}=\Pi^t_{s=1}\alpha_s$, $\epsilon_{\theta}$ is a function approximator used to predict noise from $x_t$, and $\xi\sim\mathcal{N}(\mathbf{0},\mathbf{I})$ represents uncertainty. 
This uncertainty is a source of the distribution shift. The term $\xi$ introduces randomness into the reconstruction process. Although this randomness is essential for modeling the inherent uncertainty in the data, it can lead to variability in the generated samples, causing them to deviate from the distribution of the original training data. 

We now explain how the proposed inter-class filtering can alleviate the distribution shift. Consider the distribution of the target model's training set as $\mathcal{P}_{train}$, and the distribution of the initially generated dataset as $\mathcal{P}_{gen}$. After applying inter-class filtering, the modified distribution of the generated dataset is denoted as $\mathcal{P}'_{gen}$. The Kullback-Leibler (KL) divergence, which quantifies the difference between $\mathcal{P}_{train}$ and $\mathcal{P}_{gen}$, can be mathematically formulated as follows:
\begin{equation}\label{eq:KLGen}
    KL(\mathcal{P}_{train}||\mathcal{P}_{gen})=\int\mathcal{P}_{train}(x)log(\frac{\mathcal{P}_{train}(x)}{\mathcal{P}_{gen}(x)})\mathrm{d}x.
\end{equation}
Similarly, the KL divergence between $\mathcal{P}_{train}$ and $\mathcal{P}'_{gen}$ can be expressed as:
\begin{equation}\label{eq:KLGen'}
    KL(\mathcal{P}_{train}||\mathcal{P}'_{gen})=\int\mathcal{P}_{train}(x)log(\frac{\mathcal{P}_{train}(x)}{\mathcal{P}'_{gen}(x)})\mathrm{d}x.
\end{equation}
To demonstrate the mitigation of the distribution shift, it suffices to show that $KL(\mathcal{P}_{train}||\mathcal{P}_{gen})>KL(\mathcal{P}_{train}||\mathcal{P}'_{gen})$. The detailed proof is presented below.
\begin{equation}\label{eq:KL}
\begin{aligned}
    &KL(\mathcal{P}_{train}||\mathcal{P}_{gen})-KL(\mathcal{P}_{train}||\mathcal{P}'_{gen})\\
    =&\int\mathcal{P}_{train}(x)log(\frac{\mathcal{P}_{train}(x)}{\mathcal{P}_{gen}(x)})\mathrm{d}x-\int\mathcal{P}_{train}(x)log(\frac{\mathcal{P}_{train}(x)}{\mathcal{P}'_{gen}(x)})\mathrm{d}x\\
    =&\int_{x\in\mathcal{D}_{aux}-\widehat{\mathcal{D}_{aux}}}\mathcal{P}_{train}(x)log(\frac{\mathcal{P}_{train}(x)}{\mathcal{P}_{gen}(x)})\mathrm{d}x+\\
    &\int_{x\in\widehat{\mathcal{D}_{aux}}}\mathcal{P}_{train}(x)[log(\frac{\mathcal{P}_{train}(x)}{\mathcal{P}_{gen}(x)})-log(\frac{\mathcal{P}_{train}(x)}{\mathcal{P}'_{gen}(x)})]\mathrm{d}x\\
    =&\int_{x\in\mathcal{D}_{aux}-\widehat{\mathcal{D}_{aux}}}\mathcal{P}_{train}(x)log(\frac{\mathcal{P}_{train}(x)}{\mathcal{P}_{gen}(x)})\mathrm{d}x+\\
    &\int_{x\in\widehat{\mathcal{D}_{aux}}}\mathcal{P}_{train}(x)log\frac{\mathcal{P}'_{gen}(x)}{\mathcal{P}_{gen}(x)}\mathrm{d}x\\
    \geq&\int_{x\in\widehat{\mathcal{D}_{aux}}}\mathcal{P}_{train}(x)log\frac{\mathcal{P}'_{gen}(x)}{\mathcal{P}_{gen}(x)}\mathrm{d}x
\end{aligned}
\end{equation}
In Eq. \ref{eq:KL}, the second equation arises from filtering a number of samples from the generated dataset $\mathcal{D}_{aux}$, resulting in a new dataset $\widehat{\mathcal{D}_{aux}}$. The last inequality follows from the fact that any KL divergence value is larger than or equal to $0$. To demonstrate that $\int_{x\in\widehat{\mathcal{D}_{aux}}}\mathcal{P}_{train}(x)log\frac{\mathcal{P}'_{gen}(x)}{\mathcal{P}_{gen}(x)}\mathrm{d}x>0$, it suffices to show that $\frac{\mathcal{P}'_{gen}(x)}{\mathcal{P}_{gen}(x)}>1$. According to probability theory, $\mathcal{P}_{gen}(x<\infty)=1$ and thus, $\mathcal{P}_{gen}(x<\mu+3\sigma)<1$. However, since $\mathcal{P}'_{gen}$ is derived using the three-sigma rule to filter outlying data, it follows that $\mathcal{P}'_{gen}(x<\mu+3\sigma)=1$. This implies that $\mathcal{P}'_{gen}>\mathcal{P}_{gen}$, thus concluding the proof.

\begin{figure}[ht]
\centering
	\includegraphics[scale=0.4]{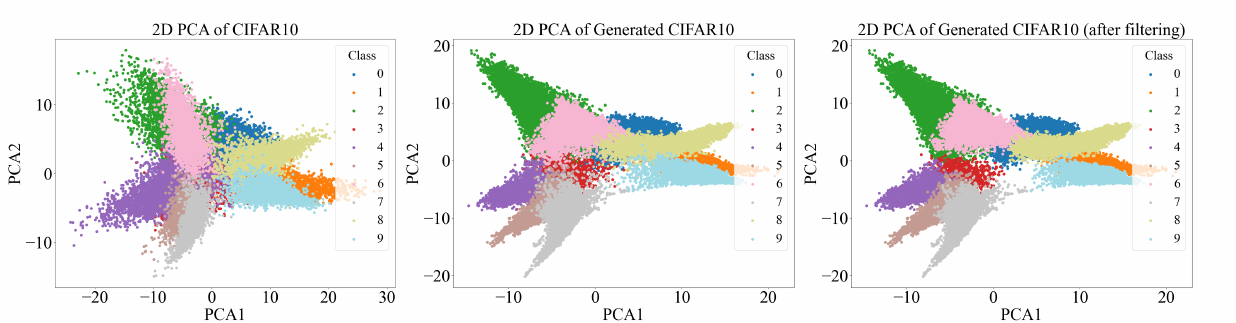}
	\caption{Data distribution of CIFAR10, generated data, and interclass-filtered data.} 
	\label{fig:plots}
\end{figure}

We also visualize the CIFAR10 data alongside the corresponding generated and interclass-filtered data. In Figure \ref{fig:plots}, the left sub-figure depicts the original CIFAR10 data distribution, the middle sub-figure shows the distribution of the generated data, and the right sub-figure illustrates the distribution after applying inter-class filtering. The inter-class filtering has effectively mitigated the distribution shift by removing outliers and refining the class boundaries in the generated dataset. This refinement is visually apparent in the right sub-figure, where the class clusters are better separated and more similar to the original CIFAR10 clusters seen in the left sub-figure. The reduction in overlap and the improved alignment of the clusters with the original data distribution support the claim that inter-class filtering helps to reduce the distribution shift from the original CIFAR10 data distribution.

\section{Experiments}
\subsection{Experimental Setup}
\noindent\textbf{Datasets.} In the experiments, we adopt three image and two language datasets. Additionally, we also created a new image dataset independently.  
\begin{itemize}[leftmargin=*]
    \item \textbf{CIFAR10} \cite{Krizhevsky14} includes $60,000$ images across $10$ classes, each containing $6,000$ images of vehicles and animals. The dimension of each image is $32\times 32$.
    \item \textbf{MNIST} \cite{LeCun98} is a dataset of $70,000$ images of handwritten numerals spanning $10$ classes: $0-9$. Each class has $7000$ images and each image was resized to $32\times 32$.
    \item \textbf{SkinCancer} \cite{SkinCancer} is a melanoma skin cancer dataset containing $10,000$ images, with $9,000$ images used for training the model and $1,000$ images for testing the model. The task involves binary classification, distinguishing between positive and negative instances of melanoma skin cancer based on skin images. Each image was resized to $32\times 32$.
    \item \textbf{BBCNews} \cite{BBCNews24} consists of RSS Feeds from the BBC News site, comprising $29,500$ records. Each record includes five attributes: title, pubDate, guid, link, and description. The dataset includes five classes: business, politics, entertainment, science, and sport.
    \item \textbf{IMDB} \cite{IMDB} contains 50,000 movie reviews categorized for binary sentiment classification, distinguishing between positive and negative sentiments. The dataset comprises 25,000 highly polar movie reviews allocated for training and an additional 25,000 for testing purposes.
\end{itemize}

In addition to utilizing the five datasets, which may have been employed to train the generative models, we have created a novel dataset named \textbf{PET}. 
The \textbf{PET} dataset was created using a combination of downloaded and self-recorded videos. Specifically, three videos each for the dog\footnote{https://www.youtube.com/shorts/8DP0TDLAt98}\footnote{https://www.youtube.com/shorts/hWlh8JHElmE}\footnote{https://www.youtube.com/watch?v=x1WSWuFqu-4} and bird\footnote{https://www.youtube.com/shorts/7a7E7VlS7c4}\footnote{https://www.youtube.com/watch?v=Dm1cblbo0S4}\footnote{https://www.youtube.com/shorts/A51r0WZxw0A} classes were downloaded from YouTube, while four videos for the cat class were recorded manually\footnote{https://github.com/orgs/SixLab6/repositories}. Images were then extracted from these videos to compile the dataset. Since the primary goal of \textbf{PET} is to develop a dataset presumed to be unseen by generative models, the number of classes is not a critical factor. However, the dataset can be easily expanded to include additional classes using the same approach.

\vspace{2mm}
\noindent\textbf{Evaluation Metrics.} As our experiments encompass various attacks, we utilize different evaluation metrics tailored to each attack's specific characteristics.

For \textbf{model extraction}, we employ \emph{accuracy} and \emph{agreement} to evaluate the efficacy of the attack. \emph{Accuracy} refers to the testing accuracy of a given test set on the stolen model. \emph{Agreement} quantifies the fraction of samples within a test set where both the target and stolen models make identical predictions.

For \textbf{membership inference}, we utilize \emph{accuracy}, \emph{F1 score}, \emph{AUC score} (area under the ROC curve), and \emph{TPR@1\%FPR} as evaluation metrics. Here, \emph{accuracy} denotes the proportion of samples within a test set whose membership status is accurately predicted by the attack model. The definitions of \emph{F1 score}, \emph{AUC score}, and \emph{TPR@1\%FPR} can be found in standard literature on classification metrics. 

For \textbf{model inversion}, we utilize \emph{MSE} (mean squared error) and \emph{accuracy} as evaluation metrics. Here, \emph{MSE} is computed between the original sample and its reconstructed counterpart. \emph{Accuracy} represents the fraction of reconstructed samples that can be correctly classified by the target model. 
In particular, \emph{MSE} is computable, as the attack is conducted in a one-to-one manner: an original sample is fed into the target model, and its output is subsequently used as input to the inversion model, which generates the reconstructed sample. Note that the original samples are used solely for evaluating attacks and are not utilized for any training purposes, as they remain inaccessible to the attacker under our threat model.

\vspace{2mm}
\noindent\textbf{Comparison Methods.} 
Comparison methods are allowed to access external data and the target model's architecture, while our method does not access such information.

For \textbf{model extraction}, the compared method assumes access to the target model's architecture and its training set. Then, a stolen model is constructed with the same architecture as the target model and trained using the target model's training set. The attack measures the output discrepancy between the stolen model and the target model as its loss. 
The key idea of this approach closely aligns with most existing attacks, such as \cite{Shen22SP,Dai23EMNLP}, which aim to optimize the stolen model's outputs to match those of the target model.

For \textbf{membership inference}, the compared method assumes knowledge of the target model's architecture and has access to its training and test sets. However, the adversary does not know which samples belong to the training or test set; otherwise, the problem would become trivial. The adversary divides the data into two parts and uses one part to train a shadow model. Data from both parts is then fed into the shadow model, and the corresponding outputs are collected. Each data point's corresponding output is labeled as either ``in'' or ``out'' based on its membership status relative to the shadow model's training set. These labeled outputs are then used to train a binary classification attack model with a cross-entropy loss.
The key idea of training shadow models to replicate the behavior of the target model and using attack models to distinguish the membership status of given samples by analyzing outputs for both members and non-members is widely adopted in existing research, such as \cite{Nasr19SP, Carlini22SP}.

For \textbf{model inversion}, the compared method assumes access to the target model's architecture and its training set. The adversary constructs a transposed CNN inversion model based on the target model's architecture. To train this inversion model, the adversary inputs the training set into the target model and collects the model's outputs. Each sample in the training set is then paired with its corresponding output. Finally, the inversion model is trained on these pairs, using the original samples as ground truth and MSE as the loss function.
The approach of constructing an inversion model against the target model and training it with the target model's outputs is broadly utilized in existing research, such as \cite{Yang19CCS,Nguyen23CVPR}.

\vspace{2mm}
\noindent\textbf{Generative Models and Prompts.} The generative models used for generating images and texts were Fast Stable Diffusion XL on TPU v5e \cite{StableDiffusion} and GPT-4.0 \cite{openai2023GPT4}, respectively. The prompts for these models were manually crafted. For instance, the prompt used for the CIFAR10 classifier was: ``Generate a single (class name), in a realistic style, with a clear background''. The size of the generated outputs were re-scaled (for images) or truncate (for texts) to fit the input dimensions of the target model. Although we explored automated prompt generation techniques, the results did not achieve the quality of those manually generated, as detailed in the ablation study. 
Moreover, details about the model architectures, the quantities of generated and augmented samples for each dataset, and corresponding computational costs are provided in the appendix. The experimental results reported are based on the average of three runs to ensure reliability in the findings.


\subsection{Overall Results}
\noindent\textbf{Model Extraction.} The experimental results, presented in Table \ref{tab:ModExt}, demonstrate that our method yields a stolen model with classification accuracy comparable to the target model, closely resembling the performance of the stolen model generated by the baseline method. Additionally, the table reveals that our method produces a stolen model with similar agreement to the target model as the one crafted by the baseline method. This indicates that both stolen models exhibit a high degree of agreement with the target model. These findings underscore the effectiveness of our approach and highlight the capability of generative models to successfully conduct model extraction attacks, despite not having access to any training data of the target model.

\begin{table}[!ht]\scriptsize
	\centering
	\caption{Model extraction results across different datasets.}
\begin{tabular} {cccc} 
\toprule
  & Target Model & \multicolumn{2}{c}{\makecell{Stolen Model\\Ours / Baseline}} \\\cline{2-4}
  & Accuracy & Accuracy & Agreement\\
 \midrule
MNIST  & $99.2$ & $98.1$ / $99.0$ & $97.8$ / $99.1$  \\
CIFAR10 & $88.7$ & $82.6$ / $85.3$ & $83.7$ / $89.4$ \\
SkinCancer & $92.3$ & $90.1$ / $90.7$ & $89.2$ / $92.8$ \\
BBCNews & $95.5$ & $87.2$ / $90.4$ & $87.7$ / $91.6$ \\
IMDB & $86.7$ & $80.6$ / $85.5$ & $86.7$ / $88.6$ \\
\bottomrule
\end{tabular}
	\label{tab:ModExt}
\end{table}

\vspace{2mm}
\noindent\textbf{Membership Inference.} Table \ref{tab:MemInf} presents the membership inference results obtained using our method and the baseline methods. Remarkably, our approach demonstrates comparable inference accuracy, F1 score, AUC and TPR@1\%FPR to the baseline method across various datasets. 
Note that we also evaluated the metric TPR@0.1\%FPR, as shown in Table \ref{tab:MemInfTPRFPR}. Achieving good results with TPR@0.1\%FPR requires training a large number of shadow models, as demonstrated in \cite{Carlini22SP}. Due to the computational cost, we limit our evaluation to TPR@1\%FPR for the remainder of this paper.

\begin{table}[!ht]\scriptsize
	\centering
	\caption{Membership inference results across different datasets.}
\begin{tabular} {ccccc} 
\toprule
  & Accuracy & F1 & AUC & TPR@1\%FPR\\\cline{2-5}
  & \multicolumn{4}{c}{Ours / Baseline}\\
 \midrule
MNIST  & $62.7$ / $61.6$ & $0.65$ / $0.63$ & $0.52$ / $0.51$ & $1.1\%$ / $1.3\%$ \\
CIFAR10 & $72.6$ / $79.4$ & $0.69$ / $0.72$ & $0.54$ / $0.67$ & $3.2\%$ / $3.5\%$ \\
SkinCancer & $69.4$ / $72.9$ & $0.61$ / $0.62$ & $0.56$ / $0.61$ & $4.9\%$ / $5.6\%$ \\
BBCNews & $70.3$ / $74.2$ & $0.59$ / $0.63$ & $0.55$ / $0.61$ & $4.4\%$ / $5.2\%$ \\
IMDB & $67.8$ / $75.8$ & $0.62$ / $0.71$ & $0.56$ / $0.67$ & $3.6\%$ / $5.2\%$ \\
\bottomrule
\end{tabular}
	\label{tab:MemInf}
\end{table}

\begin{table}[!ht]\scriptsize
	\centering
	\caption{Membership inference results on CIFAR10 with TPR@0.1\%FPR and TPR@1\%FPR.}
\begin{tabular} {cccccc} 
\toprule
 \makecell{The number of\\shadow models} & Accuracy & F1 & AUC & \makecell{TPR@\\0.1\%FPR} & \makecell{TPR@\\1\%FPR}\\
 \midrule
$1$  & $72.6$ & $0.69$ & $0.54$ & $0.4\%$ & $3.2\%$ \\
$256$ & $80.4$ & $0.77$ & $0.74$ & $2.7\%$ & $24.7\%$ \\
\bottomrule
\end{tabular}
	\label{tab:MemInfTPRFPR}
\end{table}

\vspace{2mm}
\noindent\textbf{Model Inversion.} Table \ref{tab:ModInv} illustrates the results of the inversion process conducted using both our method and the baseline method. Similar to model extraction and membership inference, our approach for model inversion also demonstrates comparable performance to the baseline method in terms of the MSE value and accuracy of the reconstructed samples.

\begin{table}[!ht]\scriptsize
	\centering
	\caption{Model inversion results across different datasets.}
\begin{tabular} {ccc} 
\toprule
  & \makecell{MSE\\Ours / Baseline} & \makecell{Accuracy\\Ours / Baseline} \\
 \midrule
MNIST  & $0.061$ / $0.042$ & $91.2$ / $95.6$  \\
CIFAR10 & $0.316$ / $0.283$ & $60.8$ / $67.1$   \\
SkinCancer & $0.174$ / $0.168$ & $78.1$ / $79.6$  \\
BBCNews & $0.627$ / $0.475$ & $60.6$ / $63.5$  \\
IMDB & $0.883$ / $0.826$ & $61.7$ / $60.8$ \\
\bottomrule
\end{tabular}
	\label{tab:ModInv}
\end{table}

\begin{figure}[ht]
\centering
	\includegraphics[scale=0.5]{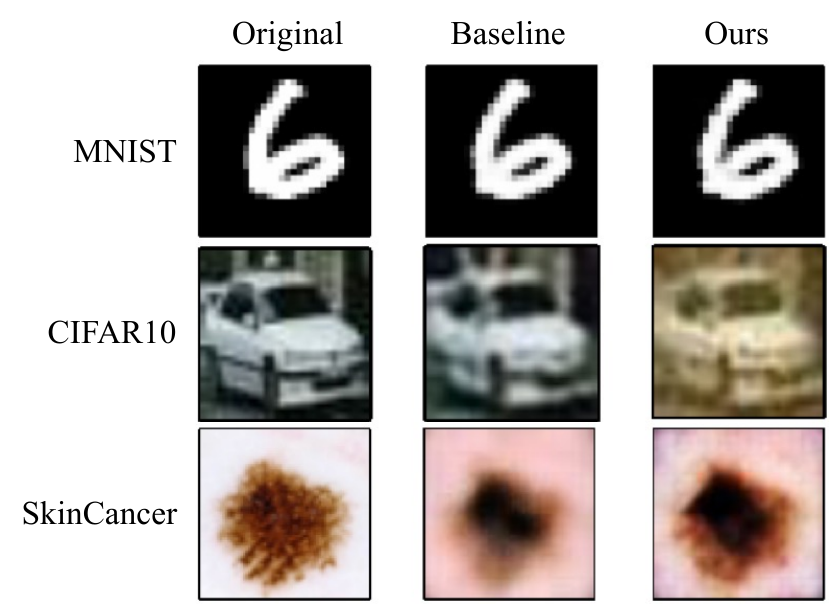}
	\caption{Inversion results on MNIST, CIFAR10 and SkinCancer.} 
	\label{fig:ImageInversion}
\end{figure}

\begin{figure}[ht]
\centering
	\includegraphics[scale=0.4]{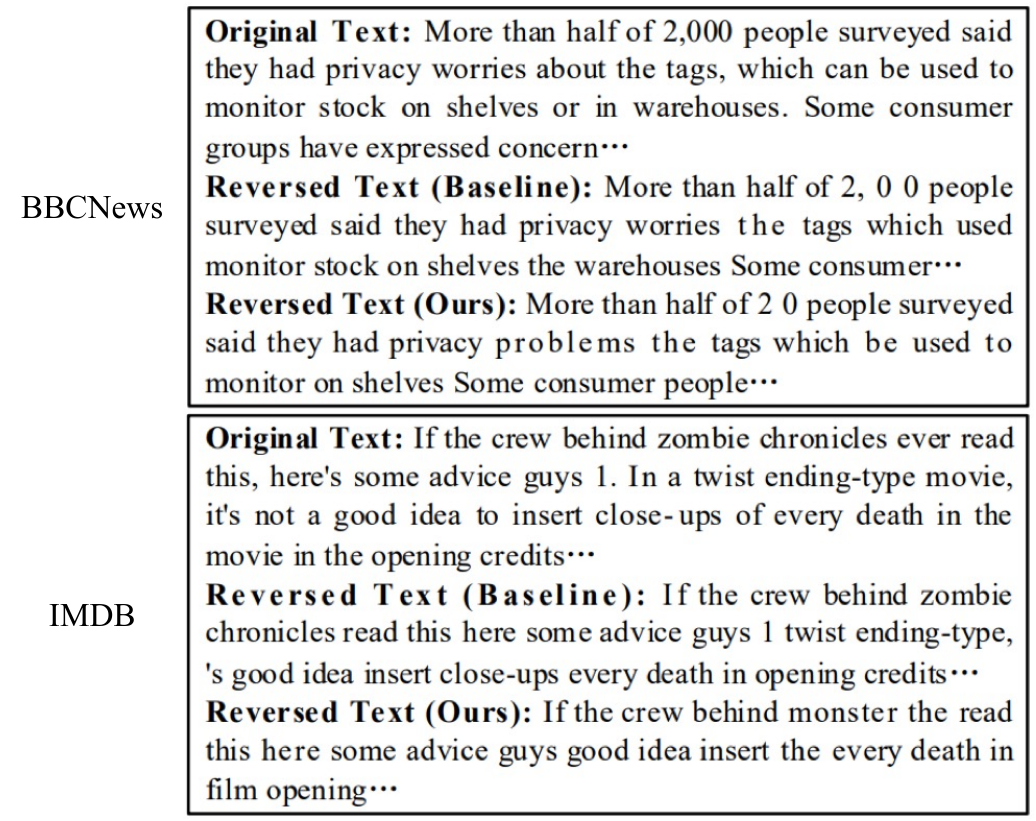}
	\caption{Inversion results on BBCNews and IMDB.} 
	\label{fig:TextInversion}
\end{figure}

We also present the model inversion results visually in Figures \ref{fig:ImageInversion} and \ref{fig:TextInversion}. It is evident that the images and texts reconstructed by our method closely resemble those reconstructed by the baseline method. 

\vspace{2mm}
\noindent\textbf{Analyses.} In the three inference attacks, a consistent pattern emerges. Both the proposed and baseline methods exhibit strong performance when the target model is trained on datasets with straightforward features, such as MNIST and SkinCancer. However, when the target model is trained on datasets with more intricate features, such as CIFAR10 and BBCNews, both methods yield relatively poorer results. This can be attributed to the higher dimensionality of features in these complex datasets, which introduce greater challenges in inferring model properties. The rich feature set in these datasets can obscure underlying patterns and relationships, making it more difficult for inference attacks to succeed.

However, there is an exception concerning membership inference, where both our method and the baseline methods exhibit deteriorative performance on MNIST. This is likely because models trained on MNIST typically achieve good generalizability to unseen data, which minimizes the differences in model outputs between member and non-member data. Consequently, this similarity in outputs reduces the effectiveness of membership inference attacks.





\vspace{2mm}
\noindent\textbf{Results for the PET dataset.} We have conducted a separate evaluation for PET, distinguishing it from others due to its private nature. This allows us to specifically demonstrate the capabilities of generative models when applied to unseen data.

\begin{table}[!ht]\scriptsize
	\centering
	\caption{Results for the PET dataset.}
\begin{tabular} {cccccccc} 
\toprule
\multicolumn{2}{c}{Model Extraction} & \multicolumn{4}{c}{Membership Inference} & \multicolumn{2}{c}{Model Inversion}\\\hline
Accu. & Agree. & Accu.& F1 & AUC & \makecell{TPR@\\1\%FPR} & MSE & Accu.\\
$86.1$ & $88.3$ & $67.6$ & $0.52$ & $0.53$ & $4.5\%$ & $0.054$ & $69$\\
\bottomrule
\end{tabular}
	\label{tab:PET}
\end{table}

\begin{figure}[ht]
\centering
	\includegraphics[scale=0.5]{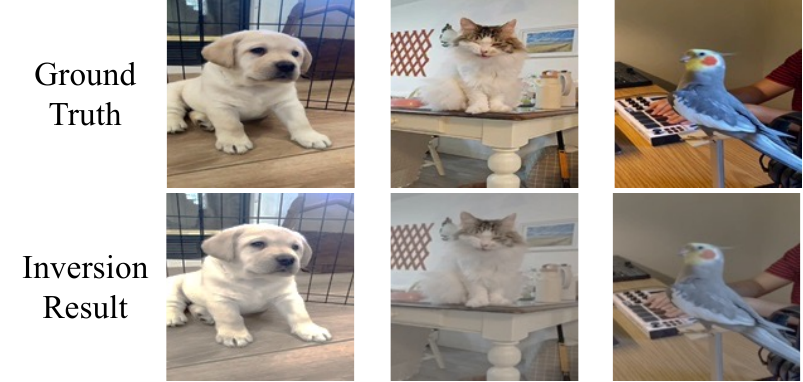}
	\caption{Inversion results on PET.} 
	\label{fig:PET}
\end{figure}

Table \ref{tab:PET} presents the numerical results of our method on the PET dataset, which align closely with those obtained from public datasets. Additionally, Figure \ref{fig:PET} visually illustrates the model inversion results on the PET dataset. These findings underscore the high generalizability of generative models, demonstrating their effectiveness in conducting model inference attacks on previously unseen datasets. 
Note that the baseline method demonstrated poor performance on the PET dataset, primarily because its effectiveness relies on the dataset size, and PET's limited size does not adequately support it. Therefore, its results on PET are not included.

\subsection{Hyperparameter Study}
Our method initially involves tasking the generative model with creating synthetic data, followed by a second step where these generated data are augmented. Thus, our hyperparameter study focuses on two aspects: the quantity of generated data and the extent of augmentation applied to these data.

\subsubsection{The Number of Generated Data} \label{subsub:dataAmount}
We assess how varying the number of generated data points for each class impacts the performance of the proposed method.
It is worth noting that due to the varying complexities of these datasets, generating samples against target models trained on them incurs different levels of computational overhead. Consequently, the number of generated samples differs for each dataset. Specifically, for MNIST, CIFAR10, and BBCNews, the number of generated samples for each class ranges from $150$ to $300$. For SkinCancer, this range extends from $200$ to $500$, while for IMDB, it spans from $50$ to $200$.

\begin{table}[!ht]\scriptsize
	\centering
	\caption{Test accuracy of the target and stolen models across different datasets with the varying number of generated data.}
\begin{tabular} {cccccc} 
\toprule
 \multirow{2}*{Accuracy} & \multirow{2}*{\makecell{Target\\Model}} & \multicolumn{4}{c}{Number of Generated Samples}\\\cline{3-6}
 & & $\scriptstyle{\textcolor{green}{50}/150/\textcolor{orange}{200}}$ & $\scriptstyle{\textcolor{green}{100}/200/\textcolor{orange}{300}}$ & $\scriptstyle{\textcolor{green}{150}/250/\textcolor{orange}{400}}$ & $\scriptstyle{\textcolor{green}{200}/300/\textcolor{orange}{500}}$ \\
 \midrule
 MNIST  & $99.2$ & $98.3$ & $99.2$ & $98.9$ & $99.4$ \\
 CIFAR10 & $88.7$ & $71.5$ & $77.2$ & $80.6$ & $82.3$ \\
 \textcolor{orange}{SkinCancer} & $92.3$ & $82.7$ & $87.8$ & $89.3$ & $90.1$ \\
BBCNews & $95.5$ & $81.9$ & $87.2$ & $89.4$ & $89.6$  \\
\textcolor{green}{IMDB} & $86.7$ & $71.3$ & $77.9$ & $79.3$ & $80.6$  \\
\bottomrule
\end{tabular}
	\label{tab:ModExtAccuData}
\end{table}

\vspace{2mm}
\noindent\textbf{Model Extraction.} The results of model extraction are shown in Table \ref{tab:ModExtAccuData}. It is evident that as the number of generated data points increases, the accuracy of the stolen models also improves across different datasets. This phenomenon underscores the efficacy of increasing the dataset with more generated data in enhancing the outcomes of model extraction. Upon closer inspection of the table, two observations emerge. First, increasing the number of generated data points has a minimal impact on simple datasets, such as MNIST.
Second, merely increasing the number of generated samples does not consistently enhance the model extraction performance. For instance, in BBCNews, while increasing the number from $150$ to $200$ leads to a $5.3\%$ improvement in accuracy, further increasing it from $250$ to $300$ results in only a $0.2\%$ increase.

The minimal impact on simple datasets, such as MNIST, can be attributed to their inherent characteristics. MNIST consists of well-separated classes and relatively straightforward patterns, facilitating easier learning for the model even with a smaller amount of data.
On the other hand, diminishing returns with an increasing number of generated samples occur due to several factors. Beyond a certain threshold, adding more generated samples may introduce redundancy or overfitting. This means that the model starts to memorize the generated data instead of learning meaningful patterns from it. As a result, the additional data may not contribute significantly to improving the model's performance.

\begin{table}[!ht]\scriptsize
	\centering
	\caption{Agreement between target and stolen models across different datasets with the varying number of generated data.}
\begin{tabular} {ccccc} 
\toprule
 \multirow{2}*{Agreement} & \multicolumn{4}{c}{Number of Generated Samples}\\\cline{2-5}
 & $\textcolor{green}{50}/150/\textcolor{orange}{200}$ & $\textcolor{green}{100}/200/\textcolor{orange}{300}$ & $\textcolor{green}{150}/250/\textcolor{orange}{400}$ & $\textcolor{green}{200}/300/\textcolor{orange}{500}$ \\
 \midrule
 MNIST  & $97.4$ & $99.0$ & $99.2$ & $99.5$ \\
 CIFAR10 & $74.2$ & $76.7$ & $80.7$ & $82.6$ \\
 \textcolor{orange}{SkinCancer} & $81.5$ & $87.3$ & $88.4$ & $89.2$ \\
BBCNews & $83.2$ & $87.7$ & $90.3$ & $90.1$  \\
\textcolor{green}{IMDB} & $73.6$ & $80.3$ & $79.5$ & $81.6$  \\
\bottomrule
\end{tabular}
	\label{tab:ModExtAgreeData}
\end{table}

A consistent pattern regarding the agreement between the target and stolen models emerges from Table \ref{tab:ModExtAgreeData}, suggesting that an increase in generated data can enhance the agreement. This phenomenon can be attributed to the fact that as the number of generated data increases, the stolen model gains access to a more comprehensive representation of the underlying data distribution captured by the target model. Thus, the stolen model is better equipped to mimic the decision boundaries and classification patterns of the target model, leading to higher agreement between them.

\vspace{2mm}
\noindent\textbf{Membership Inference.} The membership inference results are presented in Tables \ref{tab:MemInfAccuData}, \ref{tab:MemInfF1Data}, \ref{tab:MemInfAUCData} and \ref{tab:MemInfTPRData}. It can be observed that the increase of generated data has a gradually diminishing impact on membership inference across all four metrics: inference accuracy, F1 score, AUC score, and TPR@1\%FPR. This phenomenon can be attributed to several factors. Firstly, membership inference is inherently a binary classification task, distinguishing between members and non-members of the target model's training dataset. Therefore, the additional generated data, although contributing to a more comprehensive dataset, may not significantly alter the discrimination boundary between members and non-members. Secondly, the nature of membership inference relies more on identifying subtle patterns in the model's behavior rather than on the sheer volume of data. Thus, while increasing the amount of generated data may refine certain aspects of the model's behavior, it might not substantially affect its susceptibility to membership inference attacks. Lastly, the complexity of the target model and the diversity of the generated data may also play a role. If the target model is already well-trained and the generated data covers a wide range of scenarios, further increase may yield diminishing returns in terms of improving membership inference performance.

\begin{table}[!ht]\scriptsize
	\centering
	\caption{Membership inference accuracy across different datasets with the varying number of generated data.}
\begin{tabular} {ccccc} 
\toprule
 \multirow{2}*{Accuracy} & \multicolumn{4}{c}{Number of Generated Samples}\\\cline{2-5}
 & $\textcolor{green}{50}/150/\textcolor{orange}{200}$ & $\textcolor{green}{100}/200/\textcolor{orange}{300}$ & $\textcolor{green}{150}/250/\textcolor{orange}{400}$ & $\textcolor{green}{200}/300/\textcolor{orange}{500}$ \\
 \midrule
 MNIST  & $63.8$ & $62.7$ & $63.5$ & $63.9$ \\
 CIFAR10 & $55.6$ & $65.8$ & $70.6$ & $72.3$ \\
 \textcolor{orange}{SkinCancer} & $58.5$ & $61.6$ & $67.1$ & $69.4$ \\
BBCNews & $63.7$ & $70.3$ & $70.9$ & $71.2$  \\
\textcolor{green}{IMDB} & $62.1$ & $67.8$ & $69.0$ & $69.7$  \\
\bottomrule
\end{tabular}
	\label{tab:MemInfAccuData}
\end{table}

\begin{table}[!ht]\scriptsize
	\centering
	\caption{F1 score across different datasets with the varying number of generated data.}
\begin{tabular} {ccccc} 
\toprule
 \multirow{2}*{F1} & \multicolumn{4}{c}{Number of Generated Samples}\\\cline{2-5}
 & $\textcolor{green}{50}/150/\textcolor{orange}{200}$ & $\textcolor{green}{100}/200/\textcolor{orange}{300}$ & $\textcolor{green}{150}/250/\textcolor{orange}{400}$ & $\textcolor{green}{200}/300/\textcolor{orange}{500}$ \\
 \midrule
 MNIST  & $0.64$ & $0.65$ & $0.66$ & $0.66$ \\
 CIFAR10 & $0.49$ & $0.61$ & $0.69$ & $0.68$ \\
 \textcolor{orange}{SkinCancer} & $0.54$ & $0.58$ & $0.61$ & $0.60$ \\
BBCNews & $0.57$ & $0.59$ & $0.60$ & $0.62$  \\
\textcolor{green}{IMDB} & $0.58$ & $0.62$ & $0.63$ & $0.61$  \\
\bottomrule
\end{tabular}
	\label{tab:MemInfF1Data}
\end{table}

\begin{table}[!ht]\scriptsize
	\centering
	\caption{AUC score across different datasets with the varying number of generated data.}
\begin{tabular} {ccccc} 
\toprule
 \multirow{2}*{AUC} & \multicolumn{4}{c}{Number of Generated Samples}\\\cline{2-5}
 & $\textcolor{green}{50}/150/\textcolor{orange}{200}$ & $\textcolor{green}{100}/200/\textcolor{orange}{300}$ & $\textcolor{green}{150}/250/\textcolor{orange}{400}$ & $\textcolor{green}{200}/300/\textcolor{orange}{500}$ \\
 \midrule
 MNIST  & $0.50$ & $0.52$ & $0.51$ & $0.51$ \\
 CIFAR10 & $0.53$ & $0.53$ & $0.54$ & $0.55$ \\
 \textcolor{orange}{SkinCancer} & $0.52$ & $0.54$ & $0.56$ & $0.56$ \\
BBCNews & $0.54$ & $0.55$ & $0.55$ & $0.55$  \\
\textcolor{green}{IMDB} & $0.53$ & $0.56$ & $0.57$ & $0.56$  \\
\bottomrule
\end{tabular}
	\label{tab:MemInfAUCData}
\end{table}

\begin{table}[!ht]\scriptsize
	\centering
	\caption{TPR@1\%FPR across different datasets with the varying number of generated data.}
\begin{tabular} {ccccc} 
\toprule
 \multirow{2}*{AUC} & \multicolumn{4}{c}{Number of Generated Samples}\\\cline{2-5}
 & $\textcolor{green}{50}/150/\textcolor{orange}{200}$ & $\textcolor{green}{100}/200/\textcolor{orange}{300}$ & $\textcolor{green}{150}/250/\textcolor{orange}{400}$ & $\textcolor{green}{200}/300/\textcolor{orange}{500}$ \\
 \midrule
 MNIST  & $0.9\%$ & $1.2\%$ & $0.9\%$ & $1.0\%$ \\
 CIFAR10 & $2.8\%$ & $2.7\%$ & $3.2\%$ & $3.0\%$ \\
 \textcolor{orange}{SkinCancer} & $2.2\%$ & $3.5\%$ & $4.7\%$ & $4.8\%$ \\
BBCNews & $1.7\%$ & $2.4\%$ & $2.9\%$ & $3.6\%$  \\
\textcolor{green}{IMDB} & $2.2\%$ & $2.8\%$ & $4.3\%$ & $4.5\%$  \\
\bottomrule
\end{tabular}
	\label{tab:MemInfTPRData}
\end{table}

However, it is worth noting a specific observation in Table \ref{tab:MemInfAccuData}: as the number of generated data increases from $150$ to $200$, the inference accuracy against the CIFAR10 classifier notably rises from $55.6\%$ to $65.8\%$. This improvement can be attributed to several factors. Firstly, CIFAR10 is a dataset with complex features, including various objects and backgrounds. Therefore, increasing the number of generated data provides a more extensive and diverse set of examples, potentially capturing a broader range of patterns present in the target model's behavior. Additionally, the CIFAR10 dataset contains a higher degree of variability compared to simpler datasets like MNIST, requiring a larger volume of data to effectively capture its rich variability. Lastly, the mimicking model $\hat{T}$ can benefit from a larger generated dataset due to the extensive feature space, allowing for better generalization and discrimination between members and non-members of the target model's training dataset.

\vspace{2mm}
\noindent\textbf{Model Inversion.} The outcomes of model inversion are depicted in Tables \ref{tab:ModInvMSEData} and \ref{tab:ModInvAccuData}. Generally, we observe an improvement in inversion performance as the number of generated data points increases, reflected in lower MSE values and higher accuracy. Nevertheless, there are instances where this trend does not hold true. 

\begin{table}[!ht]\scriptsize
	\centering
	\caption{MSE score across different datasets with the varying number of generated data.}
\begin{tabular} {ccccc} 
\toprule
 \multirow{2}*{MSE} & \multicolumn{4}{c}{Number of Generated Samples}\\\cline{2-5}
 & $\textcolor{green}{50}/150/\textcolor{orange}{200}$ & $\textcolor{green}{100}/200/\textcolor{orange}{300}$ & $\textcolor{green}{150}/250/\textcolor{orange}{400}$ & $\textcolor{green}{200}/300/\textcolor{orange}{500}$ \\
 \midrule
 MNIST  & $0.06$ & $0.07$ & $0.05$ & $0.06$ \\
 CIFAR10 & $0.46$ & $0.41$ & $0.32$ & $0.31$ \\
 \textcolor{orange}{SkinCancer} & $0.51$ & $0.39$ & $0.32$ & $0.17$ \\
BBCNews & $0.67$ & $0.63$ & $0.61$ & $0.64$  \\
\textcolor{green}{IMDB} & $1.13$ & $0.88$ & $0.91$ & $0.85$  \\
\bottomrule
\end{tabular}
	\label{tab:ModInvMSEData}
\end{table}

\begin{table}[!ht]\scriptsize
	\centering
	\caption{Accuracy of reconstructed samples across different datasets with the varying number of generated data.}
\begin{tabular} {ccccc} 
\toprule
 \multirow{2}*{Accuracy} & \multicolumn{4}{c}{Number of Generated Samples}\\\cline{2-5}
 & $\textcolor{green}{50}/150/\textcolor{orange}{200}$ & $\textcolor{green}{100}/200/\textcolor{orange}{300}$ & $\textcolor{green}{150}/250/\textcolor{orange}{400}$ & $\textcolor{green}{200}/300/\textcolor{orange}{500}$ \\
 \midrule
 MNIST  & $91.2$ & $91.7$ & $93.4$ & $94.7$ \\
 CIFAR10 & $55.5$ & $58.2$ & $60.8$ & $62.4$ \\
 \textcolor{orange}{SkinCancer} & $62.9$ & $68.5$ & $72.2$ & $78.1$ \\
BBCNews & $60.7$ & $61.6$ & $60.3$ & $61.2$  \\
\textcolor{green}{IMDB} & $58.4$ & $61.7$ & $60.8$ & $62.3$  \\
\bottomrule
\end{tabular}
	\label{tab:ModInvAccuData}
\end{table}

In Table \ref{tab:ModInvMSEData}, we observe that the increase in the number of generated data has a limited impact on MSE for MNIST. This phenomenon can be attributed to two factors. Firstly, MNIST contains relatively simple features, making it easier for the inversion model to approximate the original samples accurately with fewer generated data points. Additionally, the nature of the inversion task itself may also play a role, as certain datasets may exhibit patterns or characteristics that are more amenable to accurate inversion, regardless of the amount of generated data used.

\begin{figure}[ht]
\centering
	\includegraphics[scale=0.5]{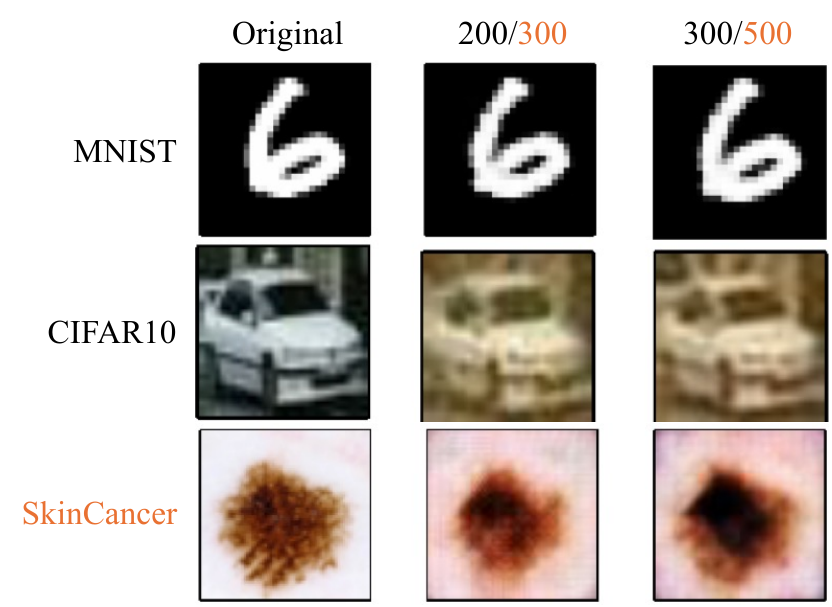}
	\caption{Inversion results on MNIST, CIFAR10 and SkinCancer with varying quantities of generated samples.} 
	\label{fig:ImageQuantity}
\end{figure}

\begin{figure}[ht]
\centering
	\includegraphics[scale=0.4]{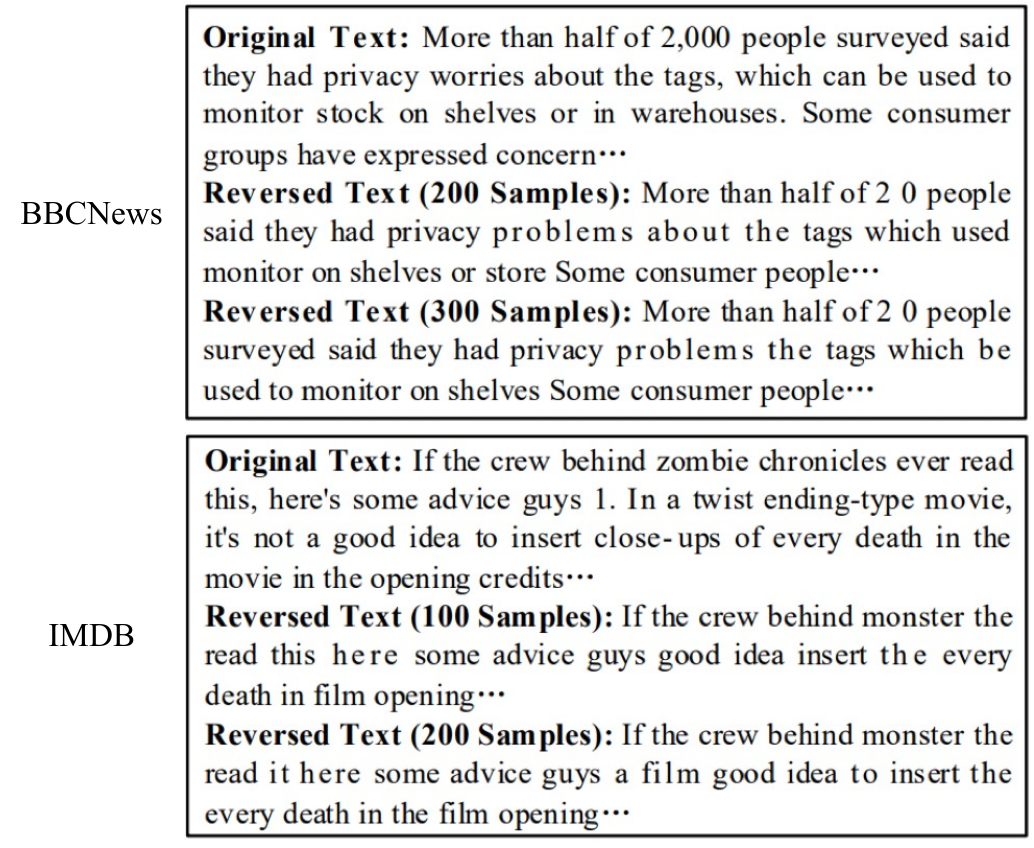}
	\caption{Inversion results on BBCNews and IMDB with varying quantities of generated samples.} 
	\label{fig:TextQuantity}
\end{figure}

We also visually present the reconstructed results in Figures \ref{fig:ImageQuantity} and \ref{fig:TextQuantity}. The reconstructed images for CIFAR10 are visually acceptable, and as the number of generated samples increases, the quality of the reconstructions improves. Similarly, for BBCNews, as the number of generated samples increases, the reconstructed texts align more closely with the original content semantically. For example, when the number of generated samples is $200$, the reconstructed text contains the phrase "which used monitor". With an increase to $300$ samples, this phrase becomes "which be used to monitor", closer to the original text. 

\subsubsection{The Impact of Augmentation}
In this setup, we assess the influence of augmenting generated data on the performance of our proposed method.

\begin{table}[!ht]\scriptsize
	\centering
	\caption{Model extraction results across different datasets with and without augmentation.}
\begin{tabular} {ccccc} 
\toprule
 \multirow{2}*{} & \multicolumn{2}{c}{With augmentation} & \multicolumn{2}{c}{Without augmentation} \\\cline{2-5}
 & Accuracy & Agreement & Accuracy & Agreement \\
 \midrule
 MNIST  & $98.1$ & $97.8$ & $95.9$ & $96.2$ \\
 CIFAR10 & $80.6$ & $80.7$ & $67.2$ & $68.0$ \\
 SkinCancer & $90.1$ & $89.2$ & $81.6$ & $80.4$ \\
BBCNews & $87.2$ & $87.7$ & $79.4$ & $79.1$  \\
IMDB & $77.9$ & $80.3$ & $72.8$ & $71.4$  \\
\bottomrule
\end{tabular}
	\label{tab:ModExtAug}
\end{table}

\vspace{2mm}
\noindent\textbf{Model Extraction.} The results of the model extraction attack with and without augmenting generated data are summarized in Table \ref{tab:ModExtAug}. It is evident that augmentation substantially enhances the performance of our method, especially noticeable in CIFAR10 where both accuracy and agreement improve by approximately $12\%$. This improvement can be attributed to several factors. First, the augmented data introduces additional diversity into the training set, allowing the stolen model to better capture the underlying patterns of the target model. Second, the augmented data may help mitigate overfitting by providing more varied samples for model training. Lastly, the augmented data could potentially address biases present in the original training data, leading to a more robust stolen model.

\begin{table*}[!ht]\scriptsize
	\centering
	\caption{Membership inference results across different datasets with and without augmentation.}
\begin{tabular} {ccccccccc} 
\toprule
 \multirow{2}*{} & \multicolumn{4}{c}{With augmentation} & \multicolumn{4}{c}{Without augmentation} \\\cline{2-9}
 & Accuracy & F1 & AUC & TPR@1\%FPR & Accuracy & F1 & AUC & TPR@1\%FPR \\
 \midrule
 MNIST  & $61.7$ & $0.65$ & $0.52$ & $1.1\%$ & $51.4$ & $0.53$ & $0.50$ & $0.7\%$ \\
 CIFAR10 & $72.6$ & $0.69$ & $0.54$ & $3.2\%$ & $33.6$ & $0.42$ & $0.49$ & $2.3\%$ \\
 SkinCancer & $69.4$ & $0.60$ & $0.56$ & $4.9\%$ & $54.1$ & $0.53$ & $0.52$ & $2.3\%$ \\
BBCNews & $70.3$ & $0.59$ & $0.55$ & $4.4\%$ & $51.7$ & $0.52$ & $0.51$ & $2.6\%$ \\
IMDB & $67.8$ & $0.60$ & $0.56$ & $3.6\%$ & $53.4$ & $0.55$ & $0.52$ & $1.9\%$ \\
\bottomrule
\end{tabular}
	\label{tab:MemInfAug}
\end{table*}

\vspace{2mm}
\noindent\textbf{Membership Inference.} The results of membership inference with and without augmenting generated data are presented in Table \ref{tab:MemInfAug}. Augmentation leads to an overall enhancement in the performance of the proposed method. 
It is observed that for CIFAR10, both the inference accuracy and F1 score improve significantly with the augmentation of generated data. Specifically, the accuracy rises from $33.6\%$ to $72.6\%$, and the F1 score increases from $0.42$ to $0.69$. This marked improvement is due to the fact that the augmentation of generated data introduces additional diversity into the training set of the attack model, thus allowing the attack model to better capture the underlying patterns of the target model's predictions.

\begin{table}[!ht]\scriptsize
	\centering
	\caption{Model inversion results across different datasets with and without augmentation.}
\begin{tabular} {ccccc} 
\toprule
 \multirow{2}*{} & \multicolumn{2}{c}{With augmentation} & \multicolumn{2}{c}{Without augmentation} \\\cline{2-5}
 & MSE & Accuracy & MSE & Accuracy \\
 \midrule
 MNIST  & $0.06$ & $91.2$ & $0.28$ & $81.6$ \\
 CIFAR10 & $0.32$ & $60.8$ & $0.52$ & $53.7$ \\
 SkinCancer & $0.17$ & $78.1$ & $0.54$ & $61.5$ \\
BBCNews & $0.63$ & $61.6$ & $1.32$ & $59.8$  \\
IMDB & $0.88$ & $61.7$ & $1.41$ & $54.2$  \\
\bottomrule
\end{tabular}
	\label{tab:ModInvAug}
\end{table}

\vspace{2mm}
\noindent\textbf{Model Inversion.} The outcomes of model inversion are shown in Table \ref{tab:ModInvAug}. The performance of the proposed method is generally improved with the augmentation of generated data, as indicated by lower MSE and higher accuracy. However, an interesting phenomenon should be noted. For BBCNews, with augmentation, the MSE reduces significantly, but the accuracy of reconstructed samples increases only marginally. This is because the augmentation of generated data introduces additional diversity and variability into the training process, allowing the inversion model to better learn the underlying patterns of the target model's predictions. This increased diversity helps to minimize the reconstruction error, leading to a reduction in MSE.
However, despite the reduction in MSE, the increase in accuracy of the reconstructed samples is relatively modest. This could be due to the inherent complexity of BBCNews, which pose challenges in accurately reconstructing the samples that share the same classification features as the original samples.

\begin{figure}[ht]
\centering
	\includegraphics[scale=0.5]{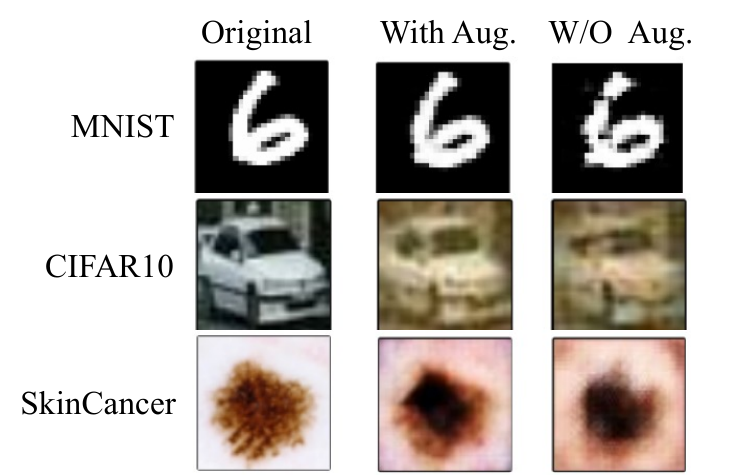}
	\caption{Inversion results on MNIST, CIFAR10 and SkinCancer with and without augmentation of generated samples.} 
	\label{fig:ImageAug}
\end{figure}

\begin{figure}[ht]
\centering
	\includegraphics[scale=0.45]{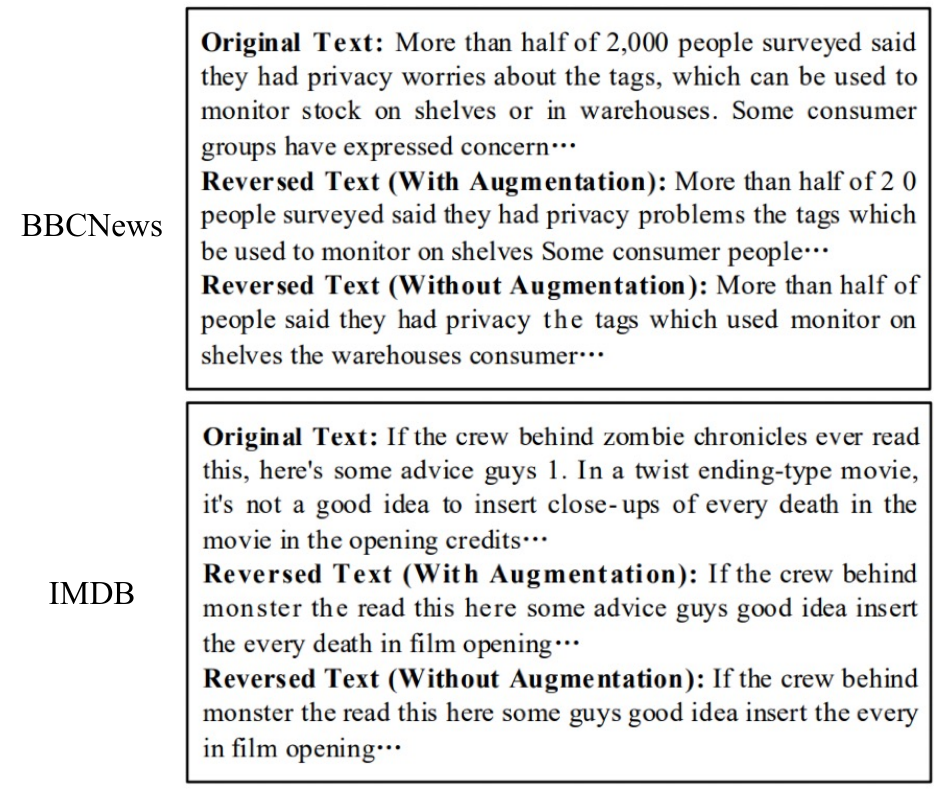}
	\caption{Inversion results on BBCNews and IMDB with and without augmentation of generated samples.} 
	\label{fig:TextAug}
\end{figure}

The visual outcomes of model inversion are depicted in Figures \ref{fig:ImageAug} and \ref{fig:TextAug}. In Figure \ref{fig:ImageAug}, it is evident that with the augmentation of generated data, the quality of reconstructed images notably improves across all three datasets: MNIST, CIFAR10, and SkinCancer, displaying higher clarity. Similarly, as illustrated in Figure \ref{fig:TextAug}, the meaning of the reconstructed texts aligns better with the original texts when augmentation is applied. For instance, in IMDB, the reconstructed text with augmentation includes two key words, `advice' and `death', which are present in the original text. Conversely, these words are not successfully reconstructed without augmentation.

\subsection{Ablation Study}
\subsubsection{Additional Baselines of Data-free Attacks}
\begin{figure}[ht]
\centering
	\includegraphics[scale=0.4]{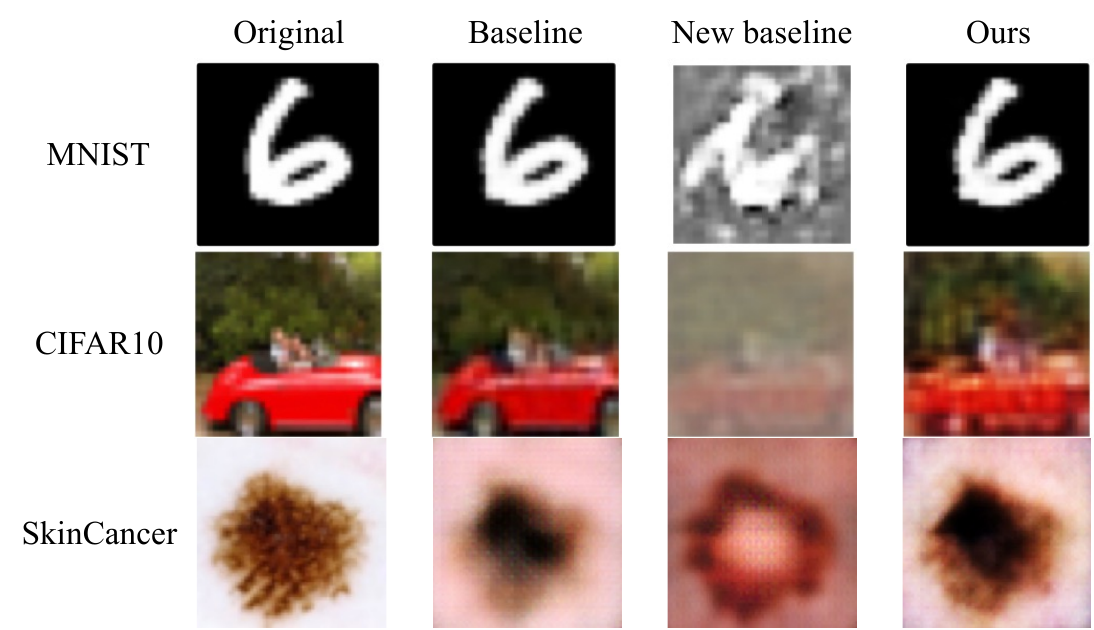}
	\caption{Inversion results on MNIST, CIFAR10 and SkinCancer using the new baseline.} 
	\label{fig:ImageNewBase}
\end{figure}

\begin{figure}[ht]
\centering
	\includegraphics[scale=0.5]{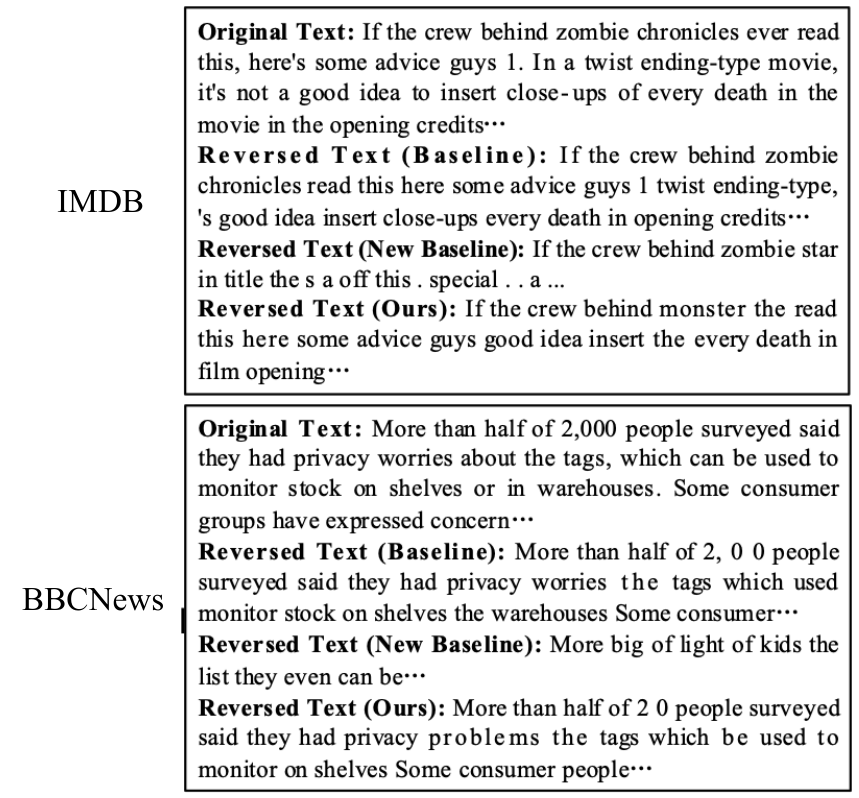}
	\caption{Inversion results on BBCNews and IMDB using the new baseline.} 
	\label{fig:TextNewBase}
    \vspace{-2mm}
\end{figure}

To evaluate the necessity of our method, we propose two additional baselines for data-free attacks: one for image models and the other for text models. Both approaches rely on searching the input space and querying the target model.
Specifically, for image target models, we generate random pixel noise, which is then structured into a noise image. This noise image is fed into the target model for classification, and the resulting classification output is used as the label for the noise image. In this manner, a dataset consisting of purely noise-based images can be generated.
Similarly, for text target models, we randomly select words and combine them into a random article. This random article is input into the target model for classification, and the classification result is used as the label for the article. This process generates a dataset of random text articles paired with their respective labels from the target model. The experimental results are presented in Tables \ref{tab:ModExtNewBase}, \ref{tab:MemInfNewBase}, \ref{tab:ModInvNewBase} and Figures \ref{fig:ImageNewBase}, \ref{fig:TextNewBase}.

It can be observed that the attack results of the new baselines are significantly inferior to those achieved by our method. This highlights the substantial advantage of using data produced by generative models over data generated through random input space exploration. This is because generative models, trained on extensive and diverse datasets, capture rich latent features and semantic relationships, enabling them to generate synthetic data that resembles real-world inputs. In contrast, random input space exploration lacks such structured knowledge and produces data that is largely uninformative, making it less effective for model-related attacks.

\vspace{-2mm}
\begin{table}[!ht]\scriptsize
	\centering
	\caption{Model extraction results across different datasets using the new baseline.}
\begin{tabular} {cccc} 
\toprule
  & Target Model & \multicolumn{2}{c}{\makecell{Stolen Model\\Ours / New baseline}} \\\cline{2-4}
  & Accuracy & Accuracy & Agreement\\
 \midrule
MNIST  & $99.2$ & $98.1$ / $60.7$ & $97.8$ / $61.5$ \\
CIFAR10 & $88.7$ & $82.6$ / $38.3$ & $83.7$ / $37.6$ \\
SkinCancer & $92.3$ & $90.1$ / $82.3$ & $89.2$ / $83.8$\\
BBCNews & $95.5$ & $87.2$ / $74.8$ & $87.7$ / $75.6$ \\
IMDB & $86.7$ & $80.6$ / $62.4$ & $86.7$ / $60.9$ \\
\bottomrule
\end{tabular}
	\label{tab:ModExtNewBase}
    \vspace{-2mm}
\end{table}

\vspace{-2mm}
\begin{table}[!ht]\scriptsize
	\centering
	\caption{Membership inference results across different datasets using the new baseline.}
\begin{tabular} {ccccc} 
\toprule
  & Accuracy & F1 & AUC & TPR@1\%FPR\\\cline{2-5}
  & \multicolumn{4}{c}{Ours / New baseline}\\
 \midrule
MNIST  & $62.7$ / $53.3$ & $0.65$ / $0.49$ & $0.52$ / $0.46$ & $1.1\%$ / $0.4\%$ \\
CIFAR10 & $72.6$ / $42.8$ & $0.69$ / $0.52$ & $0.54$ / $0.51$ & $3.2\%$ / $0.8\%$ \\
SkinCancer & $69.4$ / $57.0$ & $0.61$ / $0.58$ & $0.56$ / $0.53$ & $4.9\%$ / $2.9\%$ \\
BBCNews & $70.3$ / $60.2$ & $0.59$ / $0.55$ & $0.55$ / $0.52$ & $4.4\%$ / $1.0\%$ \\
IMDB & $67.8$ / $50.6$ & $0.62$ / $0.50$ & $0.56$ / $0.52$ & $3.6\%$ / $1.3\%$ \\
\bottomrule
\end{tabular}
	\label{tab:MemInfNewBase}
    \vspace{-2mm}
\end{table}

\vspace{-2mm}
\begin{table}[!ht]\scriptsize
	\centering
	\caption{Model inversion results across different datasets using the new baseline.}
\begin{tabular} {ccc} 
\toprule
  & \makecell{MSE\\Ours / New baseline} & \makecell{Accuracy\\Ours / New baseline} \\
 \midrule
MNIST  & $0.061$ / $1.201$ & $91.2$ / $47.6$  \\
CIFAR10 & $0.316$ / $0.572$ & $60.8$ / $28.5$   \\
SkinCancer & $0.174$ / $0.220$ & $78.1$ / $51.2$  \\
BBCNews & $0.627$ / $1.469$ & $60.6$ / $35.3$  \\
IMDB & $0.883$ / $1.611$ & $61.7$ / $32.8$ \\
\bottomrule
\end{tabular}
	\label{tab:ModInvNewBase}
\end{table}

\subsubsection{Additional Generative Models}
To evaluate the suitability of our approach, we adopted additional generative models to generate data for conducting model-related attacks. Specifically, we utilized Google Gemini \cite{Gemini} as the large language model and Open-journey \cite{Openjourney} as the image generative model. Gemini represents one of the state-of-the-art open-source large language models, while Open-journey is renowned for its efficiency and high-quality generation of high-resolution images. The experiments were conducted on CIFAR10 (for image) and IMDB (for texts). The results are presented below.

\vspace{2mm}
\noindent\textbf{Model Extraction.} The results of model extraction are presented in Table \ref{tab:ModExtOtherModels}. By comparing with Table \ref{tab:ModExt}, it is evident that using Gemini and Open-journey yields results comparable to those achieved with GPT-4.0 and Stable Diffusion. This indicates that the performance of model extraction is not strictly tied to specific generative AI models, suggesting that our approach is robust and generalizable across a variety of high-performing generative models.

\begin{table}[!ht]\scriptsize
	\centering
	\caption{Model extraction results across different datasets using additional generative models.}
\begin{tabular} {cccc} 
\toprule
  & Target Model & \multicolumn{2}{c}{\makecell{Stolen Model\\Ours / Baseline}} \\\cline{2-4}
  & Accuracy & Accuracy & Agreement\\
 \midrule
CIFAR10 & $88.7$ & $81.8$ / $85.3$ & $81.4$ / $89.4$\\
IMDB & $86.7$ & $80.2$ / $85.5$ & $83.9$ / $88.6$ \\
\bottomrule
\end{tabular}
	\label{tab:ModExtOtherModels}
    \vspace{-2mm}
\end{table}

\vspace{2mm}
\noindent\textbf{Membership Inference.} The membership inference results, presented in Table \ref{tab:MemInfOtherModels}, show a strong similarity to those in Table \ref{tab:MemInf}. This indicates that using different generative models to generate data leads to comparable membership inference results. This consistency can be attributed to the fact that popular generative models are all capable of producing synthetic data that aligns well with the underlying distribution of the target model's training data. As a result, the performance of membership inference attacks depends more on the inherent properties of the target model and the attack methodology rather than the choice of the generative model.

\begin{table}[!ht]\scriptsize
	\centering
	\caption{Membership inference results across different datasets using additional generative models.}
\begin{tabular} {ccccc} 
\toprule
  & Accuracy & F1 & AUC & TPR@1\%FPR\\\cline{2-5}
  & \multicolumn{4}{c}{Ours / Baseline}\\
 \midrule
CIFAR10 & $70.3$ / $79.4$ & $0.70$ / $0.72$ & $0.55$ / $0.67$ & $2.1\%$ / $3.5\%$ \\
IMDB & $68.5$ / $75.8$ & $0.62$ / $0.71$ & $0.57$ / $0.67$ & $4.7\%$ / $5.2\%$ \\
\bottomrule
\end{tabular}
	\label{tab:MemInfOtherModels}
\end{table}

\begin{figure}[ht]
\centering
	\includegraphics[scale=0.5]{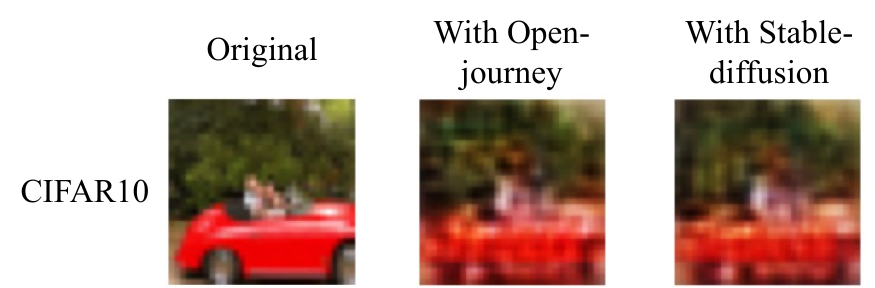}
	\caption{Inversion results on CIFAR10 with additional generative models.} 
	\label{fig:ImageOtherModels}
    \vspace{-2mm}
\end{figure}

\begin{figure}[ht]
\centering
	\includegraphics[scale=0.45]{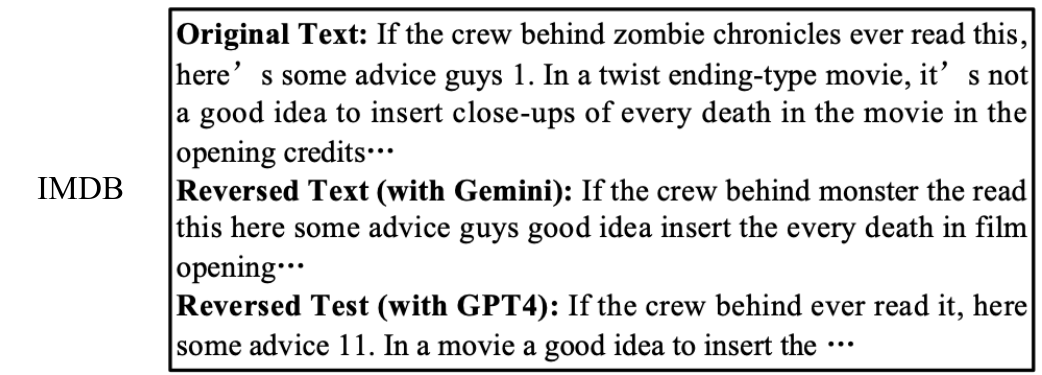}
	\caption{Inversion results on IMDB with additional generative models.} 
	\label{fig:TextOtherModels}
    \vspace{-2mm}
\end{figure}

\vspace{2mm}
\noindent\textbf{Model Inversion.} The quantitative results of model inversion are presented in Table \ref{tab:ModInvOtherModels}, which closely resemble those in Table \ref{tab:ModInv}. This indicates that current generative models are sufficiently powerful to assist adversaries in conducting model inversion attacks. This effectiveness stems from the ability of generative models to operate in a latent space that captures high-level semantic features. By leveraging these latent representations, generative models can produce outputs that align with the semantic structure of the target model’s original data, even when the specific data distribution is unknown. Additionally, the qualitative results, visually illustrated in Figures \ref{fig:ImageOtherModels} and \ref{fig:TextOtherModels}, further confirm the success of the inversion by showcasing perceptually accurate reconstructions.

\vspace{-2mm}
\begin{table}[!ht]\scriptsize
	\centering
	\caption{Model inversion results across different datasets using additional generative models.}
\begin{tabular} {ccc} 
\toprule
  & \makecell{MSE\\Ours / Baseline} & \makecell{Accuracy\\Ours / Baseline} \\
 \midrule
CIFAR10 & $0.367$ / $0.283$ & $59.6$ / $67.1$   \\
IMDB & $0.871$ / $0.826$ & $60.8$ / $60.8$ \\
\bottomrule
\end{tabular}
	\label{tab:ModInvOtherModels}
    \vspace{-2mm}
\end{table}

\subsubsection{Automated versus Manual Prompt Generation}
We have evaluated automated prompt generation techniques, such as AI Prompt Optimizer \cite{PromptPerfect}, for our attacks. Below, we provide an example comparison between a manually crafted prompt and its automated counterpart generated by the aforementioned technique.

\vspace{2mm}
\noindent\textbf{Manual prompt:} \texttt{Generate a single cat, in a realistic style, with a clear background}\\
\noindent\textbf{Automated prompt:} \texttt{Create a highly realistic image of a single cat with a clear white background. The cat should have a sleek and glossy coat, vivid green eyes, and a naturally curious expression. Ensure the fur texture is detailed and realistic, capturing the subtle variations in color and shading. The overall image should convey lifelike qualities, with attention to detail in the cat’s anatomy and fur patterns.}

\vspace{-2mm}
\begin{figure}[ht]
\centering
	\includegraphics[scale=0.38]{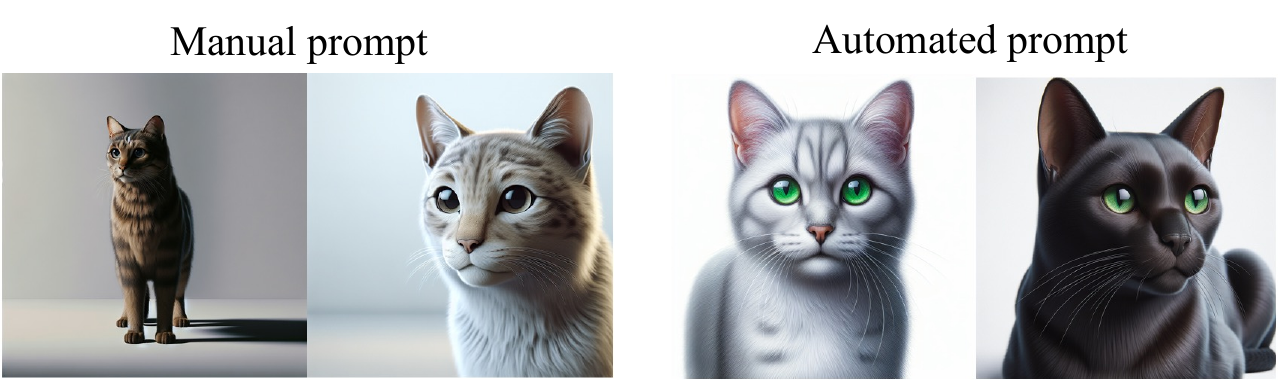}
	\caption{Comparison of images generated using manual versus automated prompts.} 
	\label{fig:PromptCompare}
    \vspace{-2mm}
\end{figure}

\vspace{2mm}
Figure \ref{fig:PromptCompare} displays the images generated from both manual and automated prompts. Visually, there are no significant perceptual differences between the images generated by these two types of prompts. To further evaluate their effectiveness, we used these images to conduct model extraction attacks against the CIFAR10 classifier by training two separate mimicking models. The results are presented in Table \ref{tab:PromptCompare}, showing that both the accuracy and agreement are comparably close.

\vspace{-2mm}
\begin{table}[!ht]\scriptsize
	\centering
	\caption{Comparison of model extraction results on CIFAR10: manual vs. automated prompts}
\begin{tabular} {cccc} 
\toprule
 \multirow{2}*{} & \multirow{2}*{Target Model} & \multicolumn{2}{c}{Stolen Model} \\\cline{3-4}
 &  &  Accuracy & Agreement \\
 \midrule
 Manual Prompt  & $88.7$ & $82.6$ & $83.7$  \\
 Automated Prompt & $88.7$ & $81.6$ & $81.4$  \\
\bottomrule
\end{tabular}
	\label{tab:PromptCompare}
    \vspace{-4mm}
\end{table}

\subsection{Summary}
The overall results highlight the potential of generative AI techniques in effectively conducting model-related attacks across various scenarios, including our newly created dataset, which is presumed to be unseen by the generative models. Notably, our proposed method achieves performance comparable to the white-box-based baseline method and significantly surpasses the random search-based baseline, further demonstrating its effectiveness and robustness.

\section{Potential Defense Strategies}

\vspace{2mm}
\noindent\textbf{Synthetic Data Detection.} As the data used in the proposed attacks are synthesized by generative models, an effective defense strategy involves detecting and filtering out synthetic data. This detection process can be accomplished by analyzing the statistical properties or patterns inherent in synthetic data. For instance, anomaly detection methods can be leveraged to identify outliers or deviations in the distribution of synthetic data compared to genuine data \cite{Gragnaniello21ICME}. 
Additionally, techniques based on feature engineering, such as examining higher-order statistical moments or texture analysis, can also help discern subtle differences between maliciously crafted synthetic data and authentic samples \cite{Yang21FGCS}. 
To specifically detect AI-generated texts, one can train a feature-based classifier capable of distinguishing between human-written and AI-generated texts using statistical measures \cite{Guo23Arxiv}, or even utilize LLMs themselves for this purpose \cite{Bhattacharjee24KDD}. 

\vspace{2mm}
\noindent\textbf{Output Perturbation.} 
As the proposed attacks rely on accessing the target model's outputs, the defender can implement intentional output perturbations to mislead the attacker. This can include various strategies such as noise injection, which adds random noise to the model's outputs to disrupt the inference process for the attacker. 
A typical technique to conduct perturbation is differential privacy \cite{Dwork14}. This involves adding carefully calibrated noise to the model's outputs to ensure that individual data points cannot be distinguished in the output, thus protecting privacy while maintaining utility.

\vspace{2mm}
\noindent\textbf{Evaluation Results.} We have evaluated one approach from each defense strategy. For synthetic data detection, we directly queried GPT-4o to identify synthetic data, successfully detecting approximately half of it. Subsequently, the evaluation is equivalent to varying the number of generated data, as demonstrated in Section \ref{subsub:dataAmount}. The results in Section \ref{subsub:dataAmount} show that our approach is not highly sensitive to the quantity of generated data. Also, to circumvent synthetic data detection, an adversary can simply request generative models to produce additional data, effectively offsetting the detection efforts.

\begin{figure}[ht]
\centering
	\includegraphics[scale=0.5]{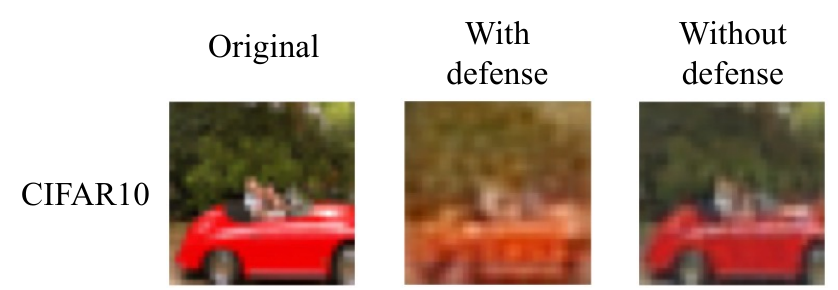}
	\caption{Inversion results on CIFAR10 with and without defense.} 
	\label{fig:ImageDefense}
\end{figure}

\begin{figure}[ht]
\centering
	\includegraphics[scale=0.4]{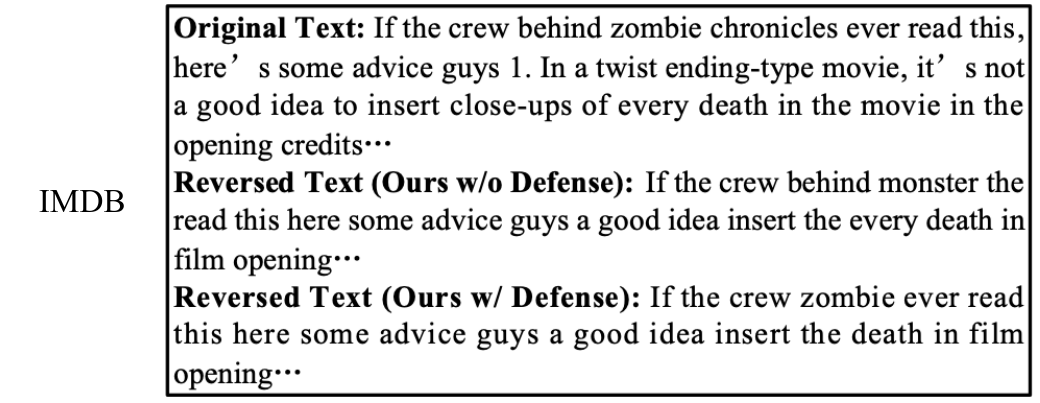}
	\caption{Inversion results on IMDB with and without defense.} 
	\label{fig:TextDefense}
\end{figure}

For output perturbation, we applied Gaussian noise with a mean of $0$ and a variance of $0.1$ to the target model's output vectors to obscure the true classification distribution. The evaluation was conducted on the CIFAR-10 and IMDB datasets, with the results presented in Tables \ref{tab:ModExtPerturbation}, \ref{tab:MemInfPerturbation}, \ref{tab:ModInvPerturbation} and Figures \ref{fig:ImageDefense}, \ref{fig:TextDefense}. These results indicate that output perturbation has moderate effectiveness in defending against model extraction attacks but has limited impact on membership inference and model inversion attacks. This is because membership inference and model inversion rely on patterns within the outputs rather than absolute precision, making them more resilient to noise introduced by perturbation.

Note that the output perturbation approach modifies each score within every output vector. While this helps obscure the true distribution, it can significantly impair the utility of the output vectors, especially in scenarios where each score carries critical importance. For example, in applications requiring detailed confidence scores for decision-making or downstream tasks, such as medical diagnostics or risk assessments, the added noise can degrade the quality of predictions, thereby reducing the overall effectiveness of the system.

\vspace{-2mm}
\begin{table}[!ht]\scriptsize
	\centering
	\caption{Model extraction results across different datasets with and without defense.}
\begin{tabular} {ccc} 
\toprule
   & Accuracy & Agreement \\\cline{2-3}
  & \multicolumn{2}{c}{With / Without Defense}\\
 \midrule
 CIFAR10 & $74.5$ / $81.6$ & $73.8$ / $82.7$  \\
IMDB & $72.4$ / $80.6$ & $73.9$ / $86.7$   \\
\bottomrule
\end{tabular}
	\label{tab:ModExtPerturbation}
    \vspace{-2mm}
\end{table}

\vspace{-2mm}
\begin{table}[!ht]\scriptsize
	\centering
	\caption{Membership inference results across different datasets with and without output defense.}
\begin{tabular} {ccccc} 
\toprule
  & Accuracy & F1 & AUC & TPR@1\%FPR\\\cline{2-5}
  & \multicolumn{4}{c}{With / Without Defense}\\
 \midrule
CIFAR10 & $65.7$ / $72.6$ & $0.61$ / $0.69$ & $0.53$ / $0.54$ & $2.2\%$ / $3.2\%$ \\
IMDB & $65.6$ / $67.8$ & $0.60$ / $0.62$ & $0.54$ / $0.56$ & $2.7\%$ / $4.4\%$ \\
\bottomrule
\end{tabular}
	\label{tab:MemInfPerturbation}
    \vspace{-2mm}
\end{table}

\vspace{-2mm}
\begin{table}[!ht]\scriptsize
	\centering
	\caption{Model inversion results across different datasets with and without output defense.}
\begin{tabular} {ccc} 
\toprule
  & MSE & Accuracy \\\cline{2-3}
  & \multicolumn{2}{c}{With / Without Defense}\\
 \midrule
CIFAR10 & $0.475$ / $0.367$ & $58.3$ / $60.8$   \\
IMDB & $0.972$ / $0.883$ & $58.6$ / $61.7$ \\
\bottomrule
\end{tabular}
	\label{tab:ModInvPerturbation}
    \vspace{-2mm}
\end{table}

\section{Related Work}

\noindent\textbf{Offensive Applications of Generative AI.} Current offensive applications of generative AI primarily focuses on cyberattacks \cite{Yao23}, such as hardware \cite{Yaman23}, OS \cite{Happe23}, software \cite{RatGPT}, network \cite{Hazell23}, and user levels \cite{Chen24ICLR}. For example, network-level attacks target vulnerabilities in network protocols, infrastructure, or communication channels. Generative AI could assist attackers in generating realistic-looking network traffic patterns, spoofed IP addresses, or social engineering messages to bypass network defenses or deceive users into revealing sensitive information. 
However, none of these attacks directly target deep learning models, especially in terms of inferring the private and sensitive information encapsulated within these models, which is the primary focus of our work.

\vspace{2mm}
\noindent\textbf{Model-Related Attacks.} 
These attacks \cite{Shen22SP,Carlini22SP,Nguyen23CVPR} typically rely on knowledge of the training data distribution of the target model. While some attacks can operate without this knowledge, their performance often suffers \cite{Kariyappa21CVPR,Mattern23ACL}. Alternatively, they may require other forms of knowledge, such as hardware- or software-based side-channels \cite{Rakin22SP}. Importantly, these attacks focus solely on individual types of attacks, whereas our method is generalized to various types of attacks.

For example, Mattern et al. \cite{Mattern23ACL} proposed a membership inference attack against language models. Their approach involves generating a set of neighboring prompts around a given prompt (e.g., by altering a word) and feeding them into the target model. By comparing the loss values output by the target model, they identify the given prompt as a member if its loss significantly differs from that of other prompts. Although their approach does not require additional information about the target model, its performance is inferior to attacks that have access to the target model's training data distribution. In contrast, our method leverages the powerful capabilities of generative AI to outperform such attack approaches without the need for additional information about the target model or the collection of external data.
Rakin et al. \cite{Rakin22SP} adapted a rowhammer memory side-channel attack to extract parameters from a CNN quantized to 8-bit. However, acquiring such side-channel information is challenging, as it often involves accessing the computing resources on which the target model is deployed \cite{Oliynyk23ACMCUR}. In contrast, our method does not require such side-channel information; instead, it relies solely on public generative models, which are much easier to access.

\section{Conclusion}
This paper has explored the utilization of generative AI techniques to facilitate model-based attacks on deep learning models. Our research shows that adversaries can leverage generative AI for model extraction, membership inference, and model inversion attacks in a data-free and black-box manner, achieving comparable performance to baseline methods that have access to target models' training data and parameters in a white-box manner.
Our findings highlight the need for robust defenses against such attacks.

Moving forward, several avenues for future research emerge. Firstly, exploring novel defense mechanisms specifically tailored to counter generative AI-based attacks is essential. Additionally, investigating the interplay between generative AI techniques and privacy-preserving measures such as differential privacy also represents an interesting direction.

\section*{Acknowledgement}
This work is partially supported by the ARC projects LP220200808 and DP230100246. This work is also funded by the European Health and Digital Executive Agency (HADEA) within the project ``Understanding the individual host response against Hepatitis D Virus to develop a personalized approach for the management of hepatitis D'' (DSolve, grant agreement number 101057917) and the BMBF with the project ``Repräsentative, synthetische Gesundheitsdaten mit starken Privatsphärengarantien'' (PriSyn, 16KISAO29K).

\section*{Ethics Considerations}
By highlighting the vulnerabilities and demonstrating the feasibility of model-related attacks using generative AI, our research seeks to underscore the potential risks associated with the deployment of such technologies. The goal is to foster a deeper understanding within the academic and tech communities about the importance of robust security measures in AI systems. To prevent any potential misuse of the methods, we will not release specific details, such as the parameters of our trained models, that could be directly exploited. 

\section*{Open Science Statement}
In this paper, we implement the baselines ourselves to provide a fair comparison with our methods. To advance research in the field of model-related attacks using generative AI, we will release our code, including the data generation, augmentation, and filtering methods, along with the newly created dataset to facilitate reproducibility and further exploration.

\bibliographystyle{plain}
\bibliography{references}

\section*{Appendix}
\setcounter{section}{0}
\renewcommand{\appendixname}{Appendix~\Alph{section}}

\section{Model Architecture, Sample Complexity, and Computation Cost}
\noindent\textbf{Model Architecture.} We utilize ResNet18 \cite{He15CVPR} for CIFAR10 classification; a custom CNN-based model with two CNN blocks and two fully-connected layers for MNIST, with an additional CNN block for PET; VGG16 \cite{Simonyan15ICLR} for SkinCancer classification; and LSTM \cite{Merity18ICLR} for both BBCNews and IMDB. 

The inversion model architectures are tailored to each dataset. For CIFAR10, the model comprises one fully connected layer followed by four pairs of CNN and transposed CNN blocks. The MNIST model includes one fully connected layer and three transposed CNN blocks, while the PET model adds an additional transposed CNN block to the MNIST structure. The SkinCancer model is designed with one fully connected layer, three transposed CNN blocks, and one additional CNN block. For textual datasets, both the BBCNews and IMDB models utilize a combination of one GRU layer and one fully connected layer.

\vspace{2mm}
\noindent\textbf{Sample Complexity.} The overall results were generated using varying sample sizes to match the complexity of each dataset: $100$ samples per class for MNIST and IMDB, $250$ samples per class for CIFAR10, $500$ samples per class for SkinCancer and PET, and $200$ samples per class for BBCNews. Additionally, the number of samples augmented for each class varied: $5,000$ for MNIST and SkinCancer; $6,000$ for CIFAR10; $2,000$ for BBCNews; $1,000$ for IMDB; and $3,000$ for PET. 

\vspace{2mm}
\noindent\textbf{Computation and Capital Cost.} The computation cost is determined by the number of generated, augmented, and filtered samples, as well as the specific generative models used. For example, generating a dataset to mimic CIFAR-10, with 250 samples per class using Fast Stable Diffusion XL on TPU v5e, required approximately 10 hours. Augmenting 3,000 samples took about 10 minutes, and filtering the same number of samples took 2 minutes.

The capital cost depends on the subscription fees for the generative AI services utilized. In our case, we subscribed to GPT-4o at a cost of \$20 USD per month, while the stable diffusion model we employed is free to use.


\end{document}